\documentclass[journal]{IEEEtran}

\usepackage{amsmath}
\usepackage{amssymb}
\usepackage{amsmath,bm}
\usepackage{enumerate}
\usepackage{graphicx}
\usepackage{subfigure}
\usepackage{cite}
\usepackage{caption}
\usepackage{algorithm}
\usepackage{algorithmic}
\usepackage{booktabs}
\usepackage{array}
\usepackage{bigstrut,bigdelim,multirow}
\usepackage{mathrsfs}
\usepackage{txfonts}
\usepackage{amsfonts,amssymb}
\usepackage{makecell}
\usepackage{epstopdf}
\usepackage{epsfig}

\begin{document}
	
	\title{MHTN: Modal-adversarial Hybrid Transfer Network for Cross-modal Retrieval}
	
	\author{Xin Huang,
		Yuxin Peng,
		and Mingkuan Yuan
		
		\thanks{This work was supported by National Natural Science Foundation of China under Grants 61371128 and 61532005.
			
			The authors are with the Institute of
			Computer Science and Technology, Peking University, Beijing 100871,
			China. Corresponding author: Yuxin Peng (e-mail: pengyuxin@pku.edu.cn).
			
		}}
		
		
		\maketitle
		$  $
		\begin{abstract}
			
			Cross-modal retrieval has drawn wide interest for retrieval across different modalities of data (such as text, image, video, audio and 3D model). However, existing methods based on deep neural network (DNN) often face the challenge of insufficient cross-modal training data, which limits the training effectiveness and easily leads to overfitting. 
			Transfer learning is usually adopted for relieving the problem of insufficient training data, but it mainly focuses on knowledge transfer only from large-scale datasets as \emph{single-modal source domain} (such as ImageNet) to \emph{single-modal target domain}.
			In fact, such large-scale single-modal datasets also contain rich modal-independent semantic knowledge that can be shared across different modalities. Besides, large-scale cross-modal datasets are very labor-consuming to collect and label, so it is significant to fully exploit the knowledge in single-modal datasets for boosting cross-modal retrieval. 
			To achieve this goal, this paper proposes modal-adversarial hybrid transfer network (MHTN), which to the best of our knowledge is the first work to realize knowledge transfer from \emph{single-modal source domain} to \emph{cross-modal target domain}, and learn cross-modal common representation. It is an end-to-end architecture with two subnetworks:
			(1) \emph{Modal-sharing knowledge transfer subnetwork} is proposed to jointly transfer knowledge from a large-scale single-modal dataset in source domain to all modalities in target domain with a star network structure, which distills modal-independent supplementary knowledge for promoting cross-modal common representation learning.
			(2) \emph{Modal-adversarial semantic learning subnetwork} is proposed to construct an adversarial training mechanism between common representation generator and modality discriminator, making the common representation \emph{discriminative for semantics} but \emph{indiscriminative for modalities} to enhance cross-modal semantic consistency during transfer process.
			Comprehensive experiments on 4 widely-used datasets show its effectiveness and generality.


		\end{abstract}
		

		%
		\IEEEpeerreviewmaketitle

		\section{Introduction}
		
\begin{figure}[t]
			\centering
			\begin{minipage}[c]{0.95\linewidth}
				\centering
				\includegraphics[width=1\textwidth]{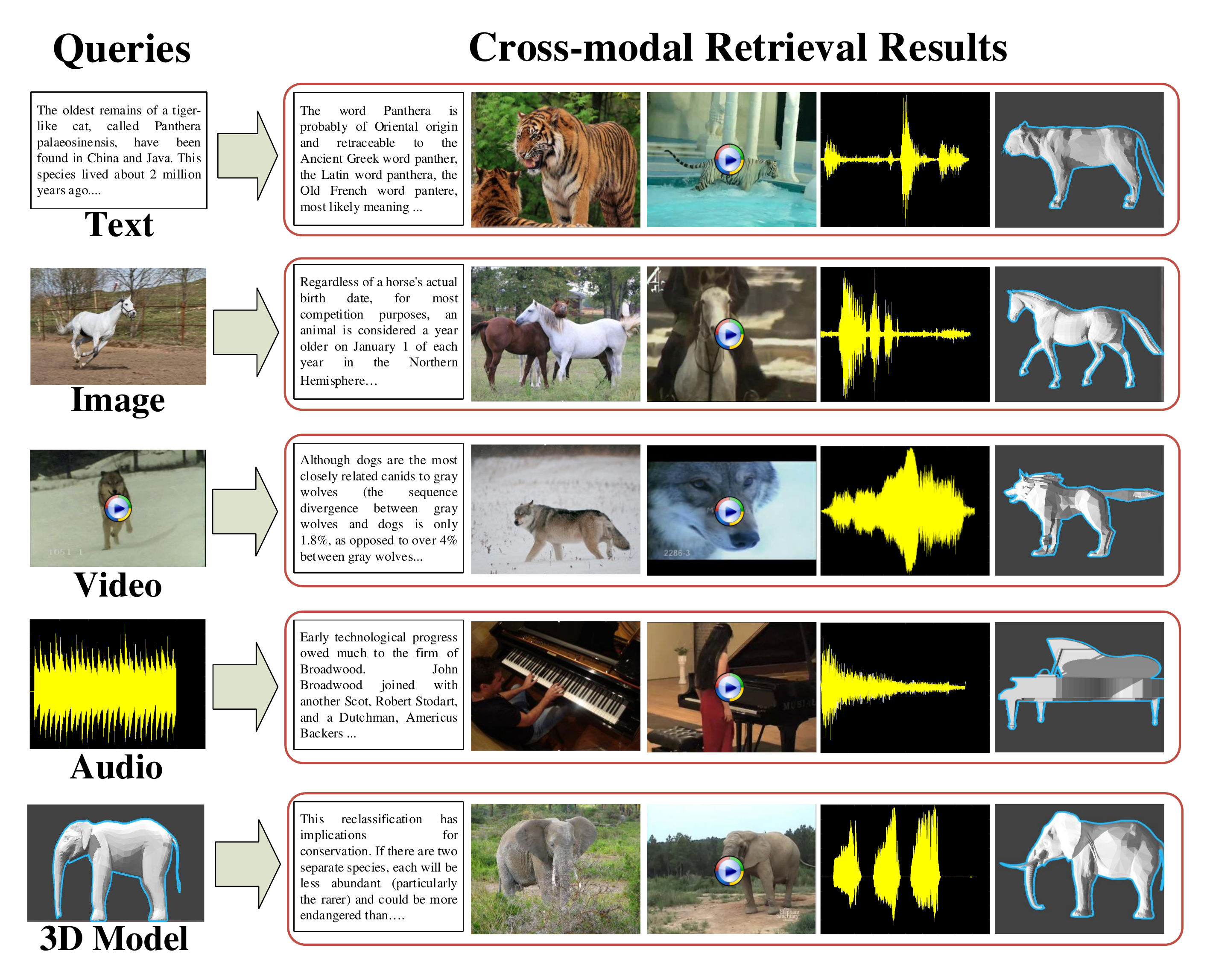}
			\end{minipage}%
			\caption{Examples of cross-modal retrieval, which can provide retrieval results with different modalities by a query of any modality.}\label{fig:example}
		\end{figure}

		In the era of big data, digital media content can be generated and found everywhere. 
		Multimodal data such as text, image, video, audio and 3D model has become the main form of information acquisition and dissemination. 
		Under this situation, cross-modal retrieval becomes a highlighted research topic, which is proposed to perform retrieval across various modalities. The essential difference between cross-modal retrieval and traditional single-modal retrieval such as \cite{PengCSVT06ClipRetrieval,DBLP:journals/tcyb/ChenXTL14}, is that cross-modal retrieval allows the modalities of query and retrieval results to be different, as shown in Figure \ref{fig:example}.
		As a novel retrieval paradigm, cross-modal retrieval has wide application prospects, and can be applied to intelligent Internet search engine and multimedia data management. However, it is also a challenging problem due to the ``heterogeneity gap", which means that the representation forms of different modalities are inconsistent, so cross-modal similarity cannot be directly computed. 
		
		For bridging ``heterogeneity gap", the existing mainstream methods follow the paradigm of common representation learning. They take the intuitive idea that there exists an intermediate common semantic space, where data of relevant semantics can be represented as similar ``feature" and be close to each other. With this idea, many methods such as \cite {HotelingBiometrika36RelationBetweenTwoVariates, RasiwasiaMM10SemanticCCA, ZhaiTCSVT2014JRL, DBLP:conf/ijcai/PengHQ16} have been proposed to learn common representation for cross-modal data, and then the cross-modal similarities can be directly computed for retrieval. Traditional methods   \cite{HotelingBiometrika36RelationBetweenTwoVariates, DBLP:journals/ijcv/GongKIL14, DBLP:conf/iccv/RanjanRJ15} mainly take linear projections as basic models. However, cross-modal correlation is highly complex to be fully captured solely by linear projections.
		Recent years, cross-modal retrieval based on deep neural network (DNN) has become an active research topic \cite{ngiam32011multimodal,srivastava2012learning,feng12014cross,DBLP:conf/ijcai/PengHQ16}. Instead of linear projections, these methods aim to fulfill the DNN's strong ability of non-linear relationship analysis in cross-modal correlation learning, and achieve accuracy improvement. However, DNN-based methods of cross-modal retrieval often face the challenge of insufficient training data, which limits the training effectiveness and easily leads to overfitting.
		
		As is known to us, insufficient training data is an important common challenge for machine learning, which is much severer for deep learning because it relies heavily on the scale of training data. For cross-modal retrieval, this problem is even greatly server. For example, if we want to collect training data for class ``tiger", we need to see the images, read the texts, watch the videos, listen to the audio, and even browse the 3D models, which is extremely labor-consuming. 
		Transfer learning \cite{DBLP:journals/tkde/PanY10, DBLP:conf/icml/LongC0J15, DBLP:conf/nips/LongZ0J16} is usually adopted for relieving the problem of insufficient training data, which transfers knowledge from large-scale datasets as source domain for boosting a specific task in target domain. But existing methods of transfer learning mainly focus on knowledge transfer only from \emph{single-modal source domain} to \emph{single-modal target domain}, and some high-quality single-modal datasets have been constructed as commonly-used source domains, such as ImageNet \cite{ImageNet2012} for image, and Google News corpus\cite{DBLP:journals/corr/abs-1301-3781} for text.

		In fact, it is significant to exploit the knowledge in these large-scale single-modal datasets for boosting cross-modal retrieval because of two reasons:
		(1) Single-modal datasets contain not only modal-specific information, but also rich modal-independent semantic knowledge that can be jointly shared across different modalities, which can provide considerable supplementary information for cross-modal retrieval. (2) Cross-modal data is much more labor-consuming to collect and label than single-modal data, and there are few large-scale labeled cross-modal datasets to serve as source domains as ImageNet. 
		So it would be of great help if the existing large-scale single-modal datasets can be fully exploited for boosting cross-modal retrieval. 
		However, it is challenging to transfer useful knowledge from \emph{single-modal source domain} to \emph{cross-modal target domain}, which is an asymmetric transfer paradigm. The knowledge from single-modal source domain cannot be directly transferred to all modalities in target domain due to ``heterogeneity gap", and the inherent cross-modal semantic consistency in target domain should be carefully preserved during the transfer process. 
		For addressing the above problems, this paper proposes modal-adversarial hybrid transfer network (MHTN), which is an end-to-end architecture with two subnetworks jointly trained to mutually boost and learn cross-modal common representation. The main contributions of this paper are as follows.
		\begin{itemize}
			\item{\emph{\textbf{Modal-sharing knowledge transfer subnetwork}}} is proposed to jointly minimize the cross-domain distribution discrepancy and cross-modal pairwise discrepancy with a star network structure, which is a hybrid transfer process from \emph{single-modal source domain} to \emph{cross-modal target domain}. Different from the existing single-modal transfer methods as \cite{DBLP:journals/JMLR/ganin2016domain, DBLP:journals/tcyb/cui2014flowing, DBLP:journals/tcyb/jiang2017integration}, this hybrid transfer structure can jointly transfer knowledge from a large-scale single-modal dataset in source domain to all modalities in target domain, distilling modal-independent supplementary information to relieve the problem of insufficient cross-modal training data.
			
			\item{\emph{\textbf{Modal-adversarial semantic learning subnetwork}}} is proposed to construct an adversarial training mechanism between \emph{common representation generator} and \emph{modality discriminator} for driving the transfer process. The former aims to generate semantic representation to be indiscriminative for modalities, while the latter tries to distinguish the modalities from the common representation, which compete each other to mutually boost. It makes the learned common representation \emph{discriminative for semantics} but \emph{indiscriminative for modalities}, thus effectively enhances cross-modal semantic consistency to improve retrieval accuracy.
			
			
		\end{itemize}

		To the best of our knowledge, MHTN is the first work for jointly transferring knowledge from \emph{single-modal source domain} to \emph{cross-modal target domain}, which can effectively transfer knowledge from large-scale single-modal datasets for addressing the problem of insufficient cross-modal training data. Compared with our previous conference paper \cite{DBLP:journals/corr/HuangPY17}, there are two major aspects of newly-added contributions achieved by this paper: 
		(1) The architecture in \cite{DBLP:journals/corr/HuangPY17} can be viewed as only a network of common representation generator. While this paper proposes a modal-adversarial training strategy with a newly-added modality discriminator for distinguishing different modalities, which competes against the common representation generator to mutually boost. This strategy can make the common representation \emph{discriminative for semantics} but \emph{indiscriminative for modalities}, which explicitly reduces ``heterogeneity gap" and enhances the cross-modal semantic consistency.
		(2) The architecture in \cite{DBLP:journals/corr/HuangPY17} has only two linked pathways in target domain, so the cross-modal retrieval is only limited to \emph{2 modalities} (image and text). While this paper further proposes a five-pathway star network structure, where the knowledge transfer and common representation learning are jointly conducted for  \emph{up to 5 modalities} (text, image, video, audio and 3D model). Because different modalities represent different aspects of semantics, jointly modeling them can allow different modalities to naturally align each other, which effectively distills the modal-independent information and boosts retrieval accuracy.
		Extensive experiments compared with 10 state-of-the-art methods on 4 widely-used cross-modal datasets show the effectiveness and generality of our MHTN approach, including the challenging XMedia dataset with up to 5 modalities.
		
		The rest of this paper is organized as follows: Section II presents a brief review of related work. Section III introduces the our proposed MHTN approach in detail. Section IV presents the comparison experiments, and Section V concludes this paper.

		\section{Related Work}
		In this section, we briefly review the works related to this paper from two aspects: cross-modal retrieval, and transfer learning. Among these, cross-modal retrieval is the research problem, and transfer learning is the main starting point for our proposed MHTN approach.
		
		\subsection{Cross-modal Retrieval}
		As introduced in Section I, cross-modal retrieval is proposed to perform retrieval task across different modalities. The key challenge of cross-modal retrieval is ``heterogeneity gap", and the mainstream of cross-modal retrieval is to represent data of different modalities with common representation. Then the cross-modal retrieval can be performed in the same common space by commonly-used distance metric, such as Euclidean distance and cosine distance.
		Existing methods can be classified into traditional methods and DNN-based methods according to the difference of basic models. 
		
		Traditional methods mainly convert cross-modal data into common representation by linear projections. For example, canonical correlation analysis (CCA) \cite{HotelingBiometrika36RelationBetweenTwoVariates} learns linear projection matrices by maximizing pairwise correlation of cross-modal data, which has many extensions as \cite{RasiwasiaMM10SemanticCCA,DBLP:conf/iccv/RanjanRJ15}.
		Cross-modal factor analysis (CFA) \cite{LiMM03CFA} directly minimizes the Frobenius norm between the common representation of pairwise data. Recently, a few works have been proposed for incorporating various information into common representation learning, such as semi-supervised and sparse regularizations \cite{ZhaiTCSVT2014JRL}, local group based priori \cite{DBLP:journals/tmm/KangXLXP15}, and semantic hierarchy \cite{DBLP:journals/tmm/HuaWLCH16}. 
		Inspired by the considerable improvement by DNN in many single-modal tasks such as image classification \cite{ImageNet2012} and object recognition \cite{DBLP:journals/tcyb/QiaoLLXW16}, researchers have made great efforts to apply DNN to cross-modal retrieval as \cite{ngiam32011multimodal,kim2012learning,DBLP:conf/ijcai/PengHQ16,DBLP:journals/tcyb/WeiZLWLZY17}. For example, Ngiam et al. \cite{ngiam32011multimodal} propose bimodal deep autoencoder, which is an extension of restricted Boltzmann machine (RBM). Data of two modalities passes through a shared code layer, in order to learn the cross-modal correlations as well as preserve the reconstruction information.
		Deep canonical correlation analysis (DCCA) \cite{ICML2013DCCA} is a non-linear extension of CCA, and can learn the complex non-linear transformations for two modalities.
		Peng et al. \cite{DBLP:conf/ijcai/PengHQ16} propose cross-media multiple deep networks (CMDN), which jointly preserves the intra-modality and inter-modality information to generate complementary separate representations, and then hierarchically combines them for improving the retrieval accuracy.
		
		However, most of the existing works as \cite{feng12014cross,DBLP:conf/ijcai/PengHQ16} only perform model training with the cross-modal datasets. Because training data plays a key role in the performance of DNN-based methods, they often face the challenge of insufficient training data, which limits the training effectiveness and easily leads to overfitting. 
		Some recent works take auxiliary single-modal datasets to directly pre-train the network component for only one modality as \cite{DBLP:journals/corr/CastrejonAVPT16,DBLP:journals/tcyb/WeiZLWLZY17}. 
		For example, ImageNet is adopted as an auxiliary dataset in the work of \cite{DBLP:journals/tcyb/WeiZLWLZY17} to pre-train a convolutional neural networks (CNN) model, and then the images in cross-modal dataset are further used to fine-tune it. 
		However, this can be viewed as knowledge transfer only from image to image, but not jointly to all modalities in target domain, which leads to inadequate transfer and limits overall retrieval accuracy. 
		
		Our MHTN is proposed for addressing the problem of jointly transferring knowledge from a single modality to multiple modalities. It can distill the modal-independent knowledge from the single-modal large-scale dataset, and exploit it to relieve the problem of insufficient cross-modal training data, thus improving the accuracy of cross-modal retrieval. 

		\subsection{Transfer Learning}
		Human can effectively exploit the learned knowledge from known tasks to promote the learning effectiveness of a new task, resulting in high generality and scalability. Inspired by this, transfer leaning \cite{DBLP:journals/tkde/PanY10} is a natural way to relieve the problem of insufficient labeled training data, which transfers knowledge from source domain to guide the model training in target domain. Generally speaking, transfer learning holds the idea that reducing the discrepancy of different domains can make source domain model work effectively in target domain \cite{DBLP:journals/JMLR/ganin2016domain, DBLP:journals/tcyb/cui2014flowing, DBLP:journals/tcyb/jiang2017integration},  and has achieved considerable success in a lot of research areas.
		The idea of transfer learning is very important for DNN-based methods as \cite{ DBLP:conf/nips/KrizhevskySH12}, whose performance usually heavily relies on the scale of training data. If there is insufficient data for a given application, transfer learning can be used for relieving this problem as \cite{DBLP:conf/icml/LongC0J15, DBLP:conf/nips/LongZ0J16}.
		
		However, most of the existing efforts on transfer learning focus on single-modal application scenarios, where the source and target domains share the same single modality (such as Image$ \rightarrow $Image). Although there are some works that involve heterogeneous domain adaptation \cite{DBLP:journals/pami/LiDXT14, DBLP:conf/cvpr/TsaiYW16}, they deal with the problem of transferring knowledge between different feature spaces of only one same modality. Beyond these, some works are proposed to transfer knowledge from one modality to another, which is still a one-to-one transfer paradigm. For example, Tang et al. \cite{DBLP:journals/tomm/tang2016generalized} propose to transfer the semantic knowledge from texts for image classification. Zhang et al. \cite{DBLP:journals/tcyb/zhang2017semi} propose an adaptation method to improve action recognition in videos by adapting knowledge conveyed from images. Some methods as \cite{DBLP:journals/tmm/YangZX15 } propose knowledge transfer between two domains with two modalities, where the modalities of the two domains should be the same. 
		Moreover, Gupta et al. \cite{DBLP:conf/cvpr/GuptaHM16} propose to use a trained model of RGB image as a pre-trained model for Depth image by knowledge distillation  \cite{ DBLP:journals/corr/HintonVD15}, which assumes that the two kinds of images have  one-to-one pixel correspondences. Cao et al. \cite{DBLP:journals/corr/CaoL016a}
		propose transitive hashing network, which learns from an auxiliary cross-modal dataset to bridge two modalities from single-modal datasets. 
		
		From the above summarization, we can see that existing works pay little attention to knowledge transfer from \emph{single-modal source domain} to \emph{cross-modal target domain}. Because cross-modal data is much more labor-consuming to collect and label than single-modal data, it is significant to exploit the knowledge in single-modal datasets for relieving the problem of insufficient cross-modal training data. Our MHTN is proposed to realize this novel transfer paradigm, which can effectively distill modal-independent supplementary knowledge to promote the model training for cross-modal tasks, such as cross-modal retrieval.
		

		\begin{figure*}[t]
			\centering
			\begin{minipage}[c]{\linewidth}
				\centering
				\includegraphics[width=0.92\textwidth]{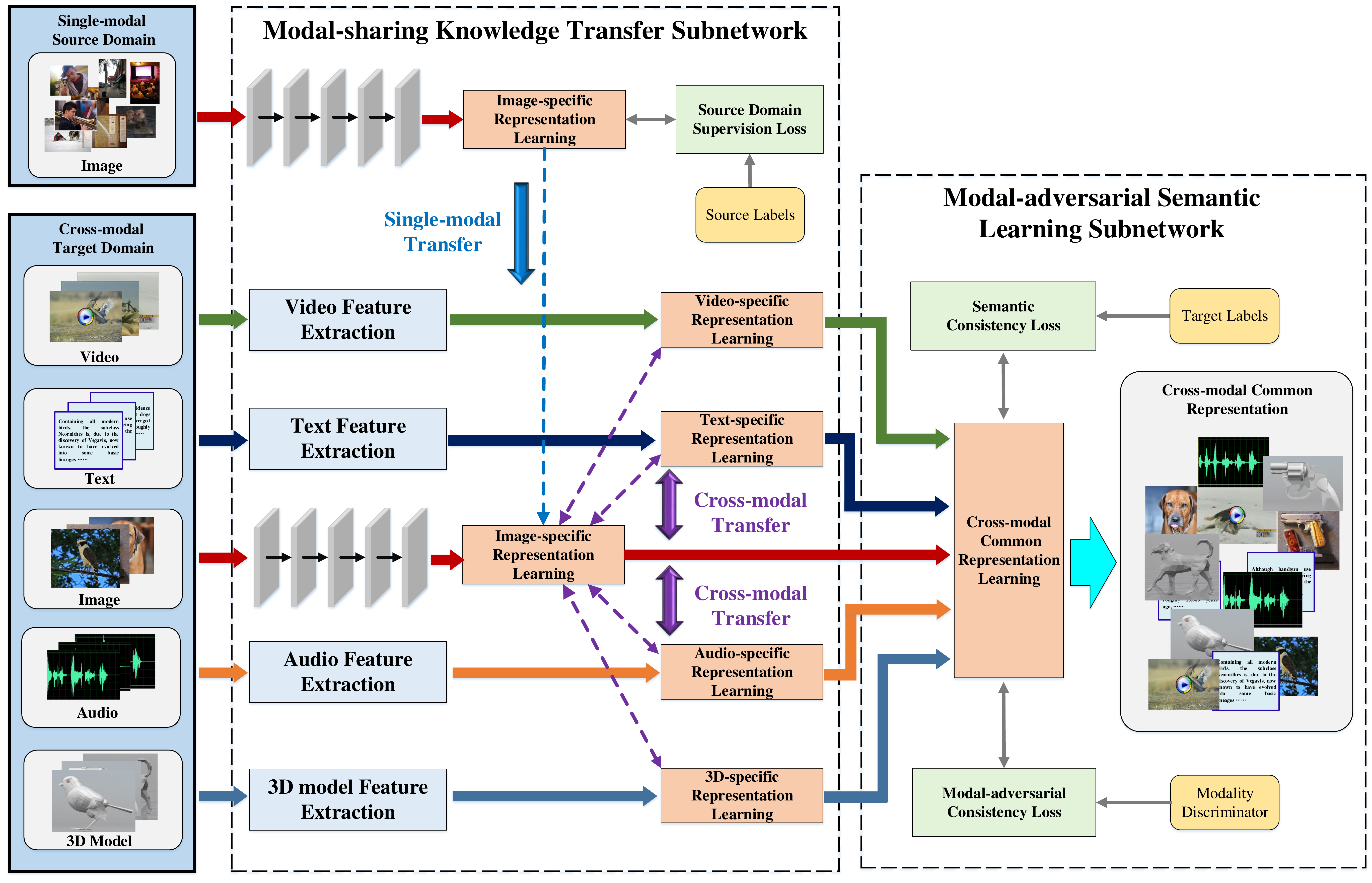}
			\end{minipage}%
			\setlength{\abovecaptionskip}{0.2cm}
			\caption{An overview of our Modal-adversarial Hybrid Transfer Network (MHTN).}\label{fig:network}
		\end{figure*}

		\section{Modal-adversarial Hybrid Transfer Network}
		
		Figure \ref{fig:network} shows the overall architecture of our MHTN approach, which takes an auxiliary image dataset as the source domain, and a cross-modal dataset with 5 modalities (text, image, video, audio and 3D model) as the target domain for example. In fact, MHTN can be also easily applied to scenarios with other number of modalities, e.g., image and text. MHTN consists of two subnetworks: (1) \emph{modal-sharing  transfer subnetwork}, and (2) \emph{modal-adversarial semantic learning subnetwork}. These two subnetworks form an end-to-end architecture, which can be trained jointly for generating cross-modal common representation.
		
		The single-modal source domain is denoted as \emph{Src}$=\{s_{i},y^{s}_{i}\}_{i=1}^{m}$ where $s_{i}$ is the $i$-th image with label $y^{s}_{i}$.
		The cross-modal target domain is denoted as \emph{Tar}$=\{D^{I}, D^{T}, D^{A} , D^{V}, D^{M}\}$, which means the data of image, text, audio, video and 3D model respectively. $D^{I} = \{ D^{I}, D^{I_u} \}$, where $D^{I} = \{d^{I}_{j},y^{I}_{j}\}_{j=1}^{N^{I}}$ denotes the labeled images for training, and $D^{I_U} = \{ d^{I_u}_{j} \}_{j=1}^{N^I_u}$ denotes the unlabeled images for testing, which is similar for $D^{T}, D^{A}, D^{V}$, and $D^{M}$. 
		For convenience, we also use term $O=\{I, T, A, V, M\}$ to denote all modalities. 
		
		The aim of the proposed MHTN is to transfer knowledge from \emph{Src} to \emph{Tar}, and train a model for generating cross-modal common representation as $R=\{R^{I}, R^{T}, R^{A}, R^{V}, R^{M}\}$ for unlabeled data. It is noted that the dimension numbers of instance in $R$ for all modalities are the same, so the cross-modal similarity can be obtained by directly computing the distance among them.


		\subsection{Modal-sharing knowledge transfer subnetwork}
		Modal-sharing knowledge transfer subnetwork is proposed to perform knowledge transfer from single-modal source domain to cross-modal target domain, which takes the modality shared by both two domains as a bridge. In training stage, we arrange the training data in \emph{Tar} as  cross-modal documents: For each image $d^I_j$, we select one instance respectively from each different modality to form a document set as $D_c = \{(d^I_j, d^T_j, d^A_j, d^V_j, d^M_j)\}_{j=1}^{N^{I}}$, where these instances are viewed with close relevance. Instances in each cross-modal document will be input into the network in parallel. If the dataset has pre-defined co-existence correlation (such as Wikipedia, NUS-WIDE-10k, and Pascal Sentences), we directly select instance according to it. Otherwise, we randomly select instances according to the semantic labels as on XMedia dataset.
		
\begin{figure}[t]
			\centering
			\begin{minipage}[c]{\linewidth}
				\centering
				\includegraphics[width=0.95\textwidth]{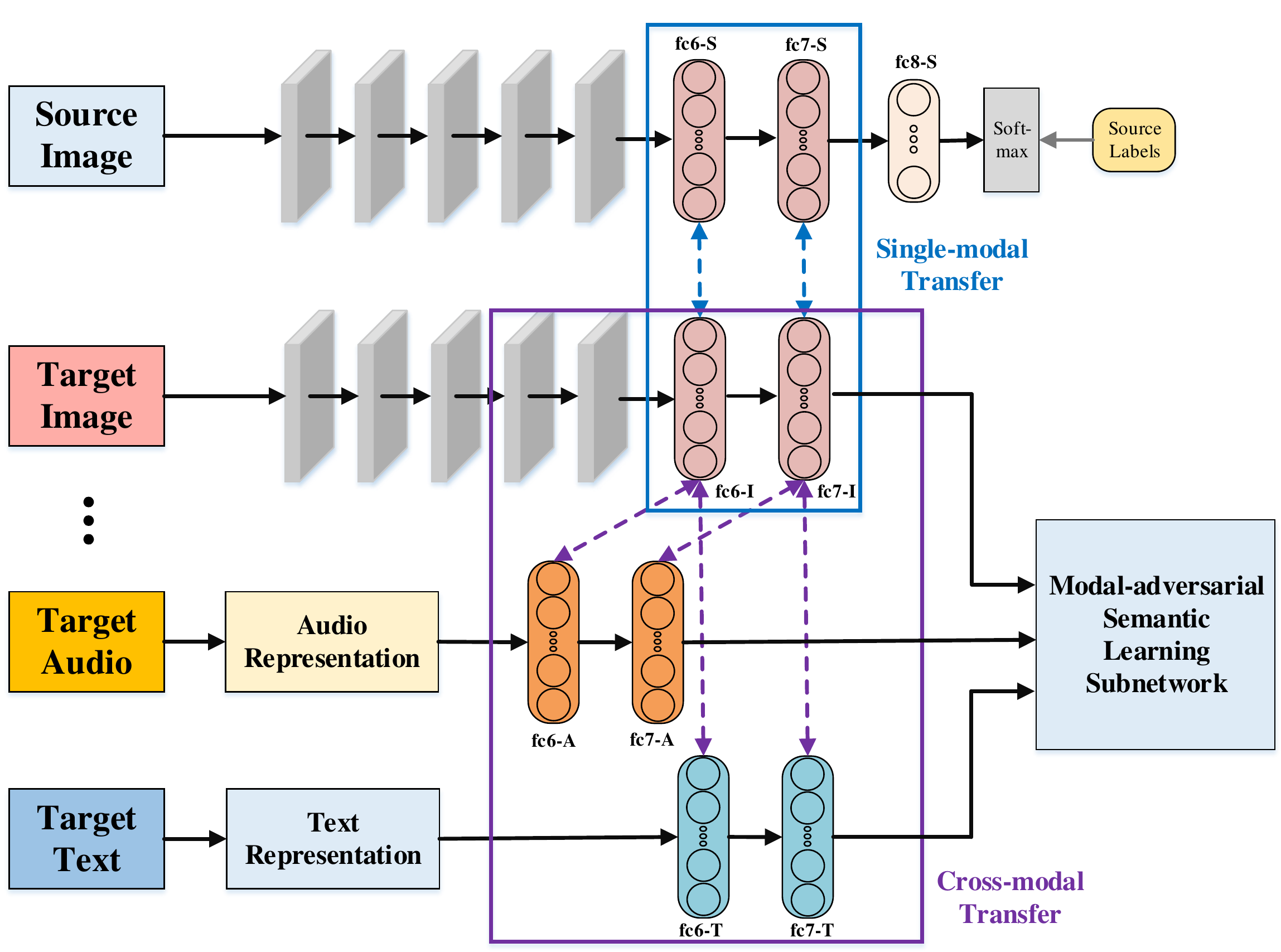}
			\end{minipage}%
			\caption{Structure of modal-sharing knowledge transfer subnetwork, where image, text and audio are shown as examples.}\label{fig:subnet1}
		\end{figure}

		Figure \ref{fig:subnet1} shows the structure of this subnetwork, where we only take three modalities (image, text and audio) as examples for clarity.  For image pathways of both source and target domains, we adopt the widely-used CNN to generate feature maps, and the two pathways are the same in the beginning of training for consistency. For the other modalities, we use extracted features as inputs. All the inputs will be fed into fully-connected layers (specific representation layers) for knowledge transfer, which is a hybrid transfer structure composed of single-modal transfer and cross-modal transfer parts.

		\subsubsection{Single-modal knowledge transfer}
		Single-modal knowledge transfer aims to allow the knowledge to be transferred from source domain to target domain through the bridge of shared modality (image). 
		We follow \cite{DBLP:conf/icml/LongC0J15} to adopt feature adaptation method \cite{gretton2012kernel} for minimizing the maximum mean discrepancy (MMD) of images between the two domains. We denote the distributions of images from source domain \emph{Src}$= \{s\} $ and target domain $D^{I} = \{d^I\}$ as $a$ and $b$ respectively, and the MMD between them is $m_k(a,b)$. We adopt the squared formulation of MMD as follows, which is in the reproducing kernel Hibert space (RKHS) $\mathcal{H}_k$:
		
		\begin{align}
			m_k^2(a,b)\overset{\Delta}{=}\left \| \mathbf{E}_a[\phi(s,\theta_S)]-\mathbf{E}_b[\phi(d^I,\theta_{I})] \right \|^2_{\mathcal{H}_k}
		\end{align}
		where $\phi$ is the representation from a network layer, and $\theta_S$ and $\theta_{I}$ denote the network parameters for \emph{Src} and $D^{I}$ respectively.  $\mathbf{E}_{\textsc{x}\sim a}f(\textsc{x})=\left \langle f(\textsc{x}),\mu _k(a) \right \rangle_{\mathcal{H}_k}$ for all $f\in \mathcal{H}_k$, where $\mu_k(a)$ is the mean embedding of $a$ in $\mathcal{H}_k$. 
		We calculate the value of $m_k^2($\emph{Src}$,D^I)$ in the corresponding specific representation layers (fc6-S/fc6-I and fc7-S/fc7-I) as the \emph{single-modal transfer loss}:
		\begin{align}
			Loss_{ST} = \sum_{l=l_6}^{l_7}m_k^2({Src},D^I)
		\end{align}
		By minimizing MMD, the model will be guided to match the distribution of target domain, so that the knowledge in source domain can be effectively transferred to target domain.
		
		Besides, we also fine-tune the source image pathway itself during the transfer process as \cite{DBLP:conf/icml/LongC0J15} with source domain data, which aims to provide supplementary supervision information for target domain, as well as preserve the semantic constraints in source domain to avoid overfitting on domain discrepancy. We define the \emph{source domain supervision loss} as follows:
		\begin{align}
			Loss_{SDS} = -\frac{1}{m}\sum_{j=1}^{m} f_{softmax}(s_j, y^{s}_j, \theta_S)
		\end{align}
		where $f_{softmax}(x, y, \theta)$ is the softmax loss function as:
		\begin{align}
			f_{softmax}(x, y, \theta) = \sum_{q=1}^c1\left \{ y=q \right \}\log \left [\hat{p}(x,q,\theta) \right ]
		\end{align}
		where $y$ denotes the label of instance $x$, $\theta$ contains network parameters and $c$ is the number of $x$'s all possible classes. If $y=q$, $1\left \{ y=q \right \}$ equals to 1, and otherwise 0.  $\hat{p}(x,q,\theta)$ is the probability distribution over classes of $x$ and can be expanded as:
		\begin{align}
			\hat{p}(x,q,\theta) = \frac{e^{\theta_q \phi(x)}}{\sum_{l=1}^c e^{\theta_l \phi(x)}}
		\end{align}
		
		By minimizing $Loss_{ST}$ and $Loss_{SDS}$, the domain discrepancy can be effectively reduced, and the supplementary semantic information from source domain can be transferred to target domain for guiding the network training.

		\subsubsection{Cross-modal knowledge transfer}
		The aforementioned part of single-modal knowledge transfer aims to allow the knowledge to be transferred between images in both two domains, but the data in source domain also contains rich and valuable modal-independent semantic knowledge that can be jointly shared across different modalities. We achieve this by cross-modal knowledge transfer, which exploits cross-modal correlation to jointly transfer knowledge to all modalities.

		Specifically, we consider the pairwise correlation between each image and instances of other modalities as \cite{LiMM03CFA,feng12014cross}. Intuitively, the network outputs of pairwise data should be similar to each other, which aims to align their representations and achieve knowledge sharing. 
		Each cross-modal document in $D_c$ can be viewed as pairs containing image and another modality. Thus each pair can be denoted as $(d^I_j, d^X_j)$, where $X\in O\land X\not= I$.
		To represent the cross-modal pairwise discrepancy, we adopt Euclidean distance between the specific representation layers of image and every other modality, which leads to a star network structure. The cross-modal pairwise discrepancy of $(d^I_j, d^X_j)$ is denoted as: 
		\begin{align}
			c^2(d^I_j,d^X_j)=\left \| \phi(d^I_j,\theta_I)-\phi(d^X_j,\theta_X) \right \|^2
		\end{align} 
		where $\phi$ is the representation from a network layer, $\theta_I$ and $\theta_X$ respectively denote the network parameters for $I$ and $X$. Then we get the \emph{cross-modal transfer loss} as:
		\begin{align}
			Loss_{CT} = \sum_{X \in O \land X \not= I}\sum_{l=l_6}^{l_7}\sum_{j=1}^{N^I}{c^2(d^I_j,d^X_j)}
		\end{align}
		By optimizing $Loss_{CT}$, the cross-modal pairwise discrepancy can be reduced to achieve cross-modal knowledge transfer.

		In the overall structure of \emph{modal-sharing knowledge transfer subnetwork}, the image modality acts as a shared bridge to link the single-modal and cross-modal transfer parts, which forms a hybrid transfer structure. By this subnetwork, the semantic knowledge contained in source domain can be jointly transferred to all modalities in cross-modal target domain. We denote the output of this subnetwork as $Z_c = \{(z^I_j, z^T_j, z^A_j, z^V_j, z^M_j)\}_{j=1}^{N^{I}}$, which will be further fed into the \emph{modal-adversarial semantic learning subnetwork}.

		\subsection{Modal-adversarial Semantic Learning Subnetwork}
		
		This subnetwork is proposed to further drive the transfer process to adapt to cross-modal retrieval task, and learn cross-modal common representation. Although the introduced \emph{modal-sharing knowledge transfer subnetwork} has actually performed hybrid knowledge transfer, there still exist two problems which limit the retrieval performance: (1) It only exploits pairwise correlation, but the high-level semantic consistency is the essential property of cross-modal target domain, which should be effectively modeled. (2) It is actually an image-centric structure, and the process of each pathway is inconsistent, which cannot fully extract the modal-independent information to reduce ``heterogeneity gap".
		
\begin{figure}[t]
			\centering
			\begin{minipage}[c]{\linewidth}
				\centering
				\includegraphics[width=0.89\textwidth]{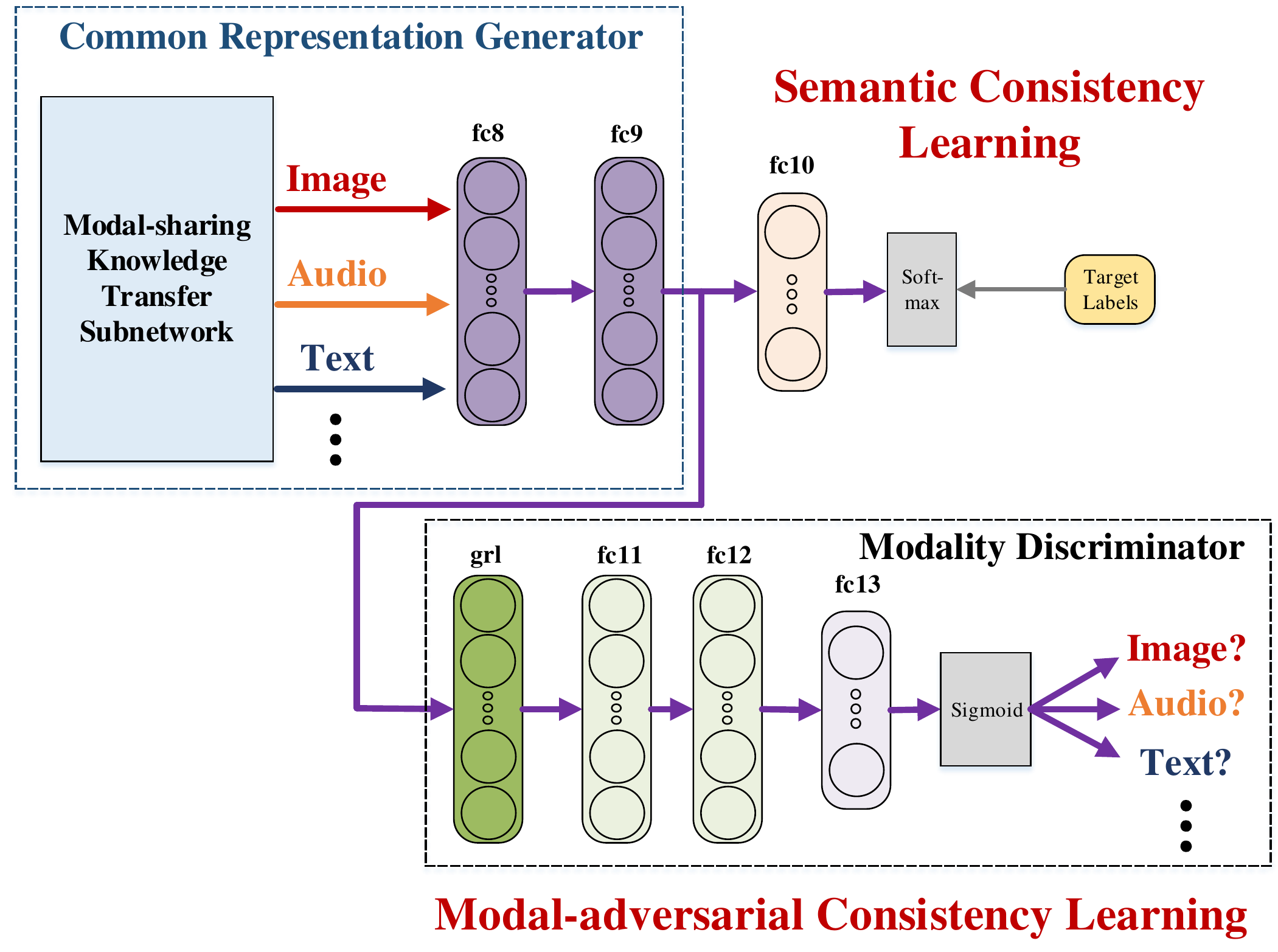}
			\end{minipage}%
			\caption{Structure of modal-adversarial semantic learning subnetwork, where image, text and audio are shown as examples.}\label{fig:subnet2}
		\end{figure}

		For addressing the above problems, we design a modal-adversarial semantic learning subnetwork. The structure of this subnetwork is shown in Figure \ref{fig:subnet2}. $Z_c$ will be fed into shared fully-connected layers (common representation layers) to generate common representation. Then there are two loss branches to drive the network training, namely \emph{semantic consistency learning}, and \emph{modal-adversarial consistency learning}.
		
		\subsubsection{Semantic consistency learning}
		In this branch, we let the cross-modal common representation be semantically discriminative. Because all modalities share the same common representation layers, the cross-modal semantic consistency can be ensured under the guidance of supervision information in the target domain.

		To achieve this goal, we use a fully-connected layer as a common classification layer with softmax loss function. The semantic consistency loss is defined as follows:
		\begin{equation}
			\begin{split}
				Loss_{SC} = -\frac{1}{N^{I}} \sum_{X \in O}\sum_{j=1}^{N^{I}} f_{softmax}(z^X_j, y_j, \theta_C)
			\end{split}
		\end{equation}
		where $z^X_j$ is the data of $X$ modality in $j$-th cross-modal document, $y_j$ is its semantic label, $\theta_C$ is the network parameter, and $f_{softmax}$ is the softmax loss function as Eq.(4).
		By optimizing $ Loss_{SC}$, we can maximize the classification accuracy jointly for all modalities, which preserves the semantic consistency contained in cross-modal target domain. 
		
		\subsubsection{Modal-adversarial consistency learning}
		Intuitively, ideal cross-modal common representation will simultaneously have two properties, both of which are very important for cross-modal retrieval: (1) It is \emph{discriminative of semantics}, so the semantic consistency of different modalities can be effectively enhanced. (2) It is \emph{indiscriminative of modalities}, so ``heterogeneity gap" is effectively reduced.
		The aforementioned branch \emph{semantic consistency learning} aims to maximize the semantic discriminative ability, while \emph{modal-adversarial consistency learning} is proposed to minimize the cross-modal representation difference.
		
		This branch can be regarded as a \emph{modality discriminator} network, while the other introduced parts of MHTN before fc10 in Figure \ref{fig:subnet2} form a \emph{common representation generator} network.  The modality discriminator aims to distinguish different modalities, while the common representation generator reduces the cross-modal representation difference to confuse the modality discriminator, which is an adversarial training style.
		
		The modality discriminator network consists of a gradient reversal layer (grl) \cite{DBLP:journals/JMLR/ganin2016domain} and fully-connected layers, the last of which is modality classification layer. The grl is an identity transform during the forward propagation, but it multiplies the gradients from the following layers by $-\lambda$ during the backpropagation, where $\lambda$ is a positive value.  In training stage, each instance is assigned with a one-hot encoding vector to indicate which modality it belongs to, and the \emph{modal-adversarial consistency loss} is:
		\begin{equation}
			\begin{split}
				Loss_{MC} = -\frac{1}{N^{I}} \sum_{X \in O} \sum_{j=1}^{N^{I}} f_{sigmoid}(z^X_j, p{(z^X_j)}, \theta_M)
			\end{split}
		\end{equation}
		where $p{(\cdot)}$ denotes the label indicator, $\theta_M$ denotes the network parameters, and $f_{sigmoid}(x, p, \theta)$ is the sigmoid cross entropy loss function following \cite{DBLP:journals/JMLR/ganin2016domain}:
		\begin{equation}
			\begin{split}
				f_{sigmoid}(x, p, \theta)	&= p(x) \log \hat{p}(x,\theta)\\
				&+ [1-p(x)] \log [1-\hat{p}(x,\theta)]
			\end{split}
		\end{equation}
		where the $\hat{p}(x,\theta)$ is as:
		\begin{align}
			\hat{p}(x,\theta) = \frac{e^{\theta \phi(x)}}{\sum_{l=1}^c e^{\theta_l \phi(x)}}
		\end{align}
		Due to the existence of grl, the gradient of this part will be reversed during the training stage. 
		By maximizing $Loss_{MC}$, we can explicitly reduce ``heterogeneity gap" among modalities and enhance the consistency of common representation.

		Our MHTN is an end-to-end architecture with \emph{modal-sharing knowledge transfer subnetwork} and \emph{modal-adversarial semantic learning subnetwork}. So the two subnetworks are trained jointly and boost each other. In testing stage, each testing instance in \emph{Tar} can be converted into predicted class probability vector as the final common representation $R$ for retrieval as \cite{DBLP:journals/tcyb/WeiZLWLZY17,ZhaiTCSVT2014JRL} . It is noted that the testing data can be input separately, which is unnecessary to be input in the form of cross-modal document.

		\subsection{Optimization}
		
		As for the network optimization, because the aforementioned loss functions are calculated in different positions of network, we should first denote the parameters of different parts for clarity: 
		In \emph{modal-sharing knowledge transfer subnetwork}, we denote the parameters of source domain pathway as $\theta_S$, the parameters of target domain pathway for image as $\theta_I$, and for all the other modalities as $\theta_{O'} = \{ \theta_T, \theta_A, \theta_V, \theta_M\}$. In \emph{modal-adversarial semantic learning subnetwork}, the parameters of modal-adversarial consistency learning part are as $\theta_M$, and the others are as $\theta_C$.
		
		With the above notations, we can assign the parameters to each loss function, and formally consider the loss function:
		\begin{equation}
			\begin{split}
				& E(\theta_S, \theta_I, \theta_{O'}, \theta_C, \theta_M)\\
				& = Loss_{ST}(\theta_S, \theta_I) + Loss_{SDS}(\theta_S) + Loss_{CT}(\theta_I, \theta_{O'}) \\
				& + Loss_{SC}(\theta_I, \theta_{O'}, \theta_C) - \lambda Loss_{MC}(\theta_I, \theta_{O'}, \theta_C, \theta_M)
			\end{split}
		\end{equation}
		where $\lambda$ is a positive trade-off parameter between the positive and negative loss functions in training stage. Our goal is to find the parameters $\theta_S$, $\theta_I$, $\theta_{O'}$, $\theta_C$, $\theta_M$ for getting the saddle point of Eq.(12):
		\begin{align}
			(\hat{\theta}_S, \hat{\theta}_I, \hat{\theta}_{O'}, \hat{\theta}_C) = \arg \min_{\theta_S, \theta_I, \theta_{O'}, \theta_C}E(\theta_S, \theta_I, \theta_{O'}, \theta_C, \hat{\theta}_M)
		\end{align}
		\begin{align}
			\hat{\theta}_M = \arg \max_{\theta_M}E(\hat{\theta_S}, \hat{\theta_I},  \hat{\theta}_{O'}, \hat{\theta_C}, \theta_M)
		\end{align}
		At the saddle point, the parameters $\theta_S$, $\theta_I$, $\theta_{O'}$, $\theta_C$ of the previous networks minimize Eq.(12), while the parameters $\theta_M$ maximize Eq.(12), which is an adversarial training style.
		Based on Eq.(13-14), we can update the parameters as follows:
		
		\begin{align}
			\theta_S \, \, \, \, \, \leftarrow \, \, \, \, \, \theta_S - \mu \Bigg ( \frac{\partial Loss_{ST}}{\partial \theta_S} + \frac{\partial Loss_{SDS}}{\partial \theta_S} \Bigg )
		\end{align}
		
		\begin{equation}
			\begin{split}
				\theta_I \, \, \, \, \, \leftarrow \, \, \, \, \,   \theta_I  & - \mu \Bigg (  \frac{\partial Loss_{ST}}{\partial \theta_I} + \frac{\partial Loss_{CT}}{\partial \theta_I} \\
				& + \frac{\partial Loss_{SC}}{\partial \theta_I} - \lambda \frac{\partial Loss_{MC}}{\partial \theta_I}  \Bigg )
			\end{split}
		\end{equation}
		
		\begin{equation}
			\begin{split}
				\theta_{O'} \, \, \, \, \, \leftarrow \, \, \, \, \,  \theta_{O'}  - \mu \Bigg (  \frac{\partial Loss_{CT}}{\partial \theta_{O'}} + \frac{\partial Loss_{SC}}{\partial \theta_{O'}} 
				- \lambda \frac{\partial Loss_{MC}}{\partial \theta_{O'}} \Bigg )
			\end{split}
		\end{equation}
		
		\begin{align}
			\theta_C \, \, \, \, \, \leftarrow \, \, \, \, \, \theta_C - \mu \Bigg ( \frac{\partial Loss_{SC}}{\partial \theta_C} - \lambda \frac{\partial Loss_{MC}}{\partial \theta_C} \Bigg )
		\end{align}
		
		\begin{align}
			\theta_M \, \, \, \, \, \leftarrow \, \, \, \, \, \theta_M - \mu \frac{\partial Loss_{MC}}{\partial \theta_M}
		\end{align}
		where $\mu$ denotes the learning rate. The parameter updates of Eq.(15-19) can be realized by stochastic gradient descent (SGD) algorithm. In this way, we can optimize these loss functions and perform knowledge transfer from single-modal source domain to cross-modal target domain in training stage, so as to get effective cross-modal common representation $R$.

		\section{Experiments}
		
		In this section, we present the experiments for verifying the effectiveness of our proposed MHTN. The implementation details of deep architecture are first introduced, and then we discuss the adopted datasets, evaluation metrics and compared methods. Next, experimental results compared with state-of-the-art methods along with the analysis are presented. Finally, we introduce the experiments on components of our MHTN to show their impacts.
		
		\subsection{Details of the Deep Architecture}
		In this section, we present the details of our MHTN. The implementation of MHTN is based on Caffe\footnote{http://caffe.berkeleyvision.org}, a widely-used deep learning framework. It is noted that the presented architecture is for 5 modalities, which can be easily applied to other number of modalities by adjusting the pathway number.
		
		\subsubsection{Modal-sharing knowledge transfer subnetwork} In the source image pathway and target image pathway, we adopt convolutional layers (conv1-conv5) of AlexNet \cite{DBLP:conf/nips/KrizhevskySH12} pre-trained on ImageNet with Caffe Model Zoo. The input images are first resized as $256 \times 256$ and then used to generate convolutional feature maps (pool5). These convolution layers are frozen in training stage, because they are regarded as general layers. All the six pathways have two fully-connected layers (specific representation layers), and all of them have $4,096$ units. The learning rates of all the fully-connected layers are fixed to be 0.01.
		
		The single-modal transfer loss is implemented by MMD loss layers \cite{DBLP:conf/icml/LongC0J15} between the source image pathway and target image pathway, while the cross-modal transfer loss is implemented by contrastive loss layers from Caffe. After fc8-S layer and a softmax loss layer of source image pathway, we can calculate $Loss_{SDS}$ for images of source domain in training stage. Moreover, because the magnitude of $Loss_{CT}$ is much larger than those of $Loss_{ST}$ and $Loss_{SDS}$ (about 1,000 times), we set its weight as 0.001, and those of  $Loss_{ST}$ and $Loss_{SDS}$ are 1. The weight decay of this subnetwork is set as 0.0005. These parameter settings can be easily adjusted in the implementation of network.
		
		\subsubsection{Modal-adversarial semantic learning subnetwork} This subnetwork has two fully-connected layers with $4,096$ units (common representation layers), which are fc8 and fc9 in Figure \ref{fig:subnet2}, and the learning rates are fixed to be 0.01. Then after a fully-connected classification layer (fc10) and a softmax layer, we can calculate $Loss_{SC}$ for all modalities of target domain in training stage, and get the probability vector as final common representation in testing stage. Besides, for the part of modal-adversarial consistency learning, there is a gradient reversal layer \cite{DBLP:journals/JMLR/ganin2016domain}, three fully-connected layers (fc11 and fc12 with 1,024 units, and fc13 with 5 units), and a sigmoid cross entropy loss layer. Note that the unit number of fc13 is the same with modality number. The learning rates of fc11-13 are fixed as 0.001. The weight decay of this subnetwork is also set as 0.0005. The weight of $Loss_{MC}$ ($\lambda$) is set as 0.1 to avoid excessive influence, while that of $Loss_{SC}$ is 1.
		
		\subsection{Datasets}
		
		In our experiments, ImageNet \cite{ImageNet2012} serves as the source domain, which is from ImageNet large-scale visual recognition challenge (ILSVRC)
		2012. Cross-modal retrieval is conducted on 4 widely-used datasets, namely Wikipedia, NUS-WIDE-10k, Pascal Sentences and XMedia datasets. Note that Wikipedia, NUS-WIDE-10k and Pascal Sentences datasets involve 2 modalities (text and image), while there are up to 5 modalities in XMedia dataset (text, image, video, audio, and 3D model). The dataset splits of Wikipedia, NUS-WIDE-10k and Pascal Sentences are strictly according to  \cite{feng12014cross,DBLP:conf/ijcai/PengHQ16}, and similar split is applied to XMedia for fair comparison. These 4 datasets are briefly introduced as follows:
		
		{\textbf {Wikipedia dataset}} \cite{RasiwasiaMM10SemanticCCA} is constructed from the ``featured articles" of Wikipedia, which is the most popular cross-modal dataset for evaluation  \cite{ZhaiTCSVT2014JRL,feng12014cross,DBLP:conf/ijcai/PengHQ16}. ``Featured articles" contain 29 classes, and the 10 most populated ones form the final dataset with 2,866 image/text pairs, where each pair only belongs to one class. In each pair, the text is several paragraphs as descriptions of the image.
		The classes are of high-level semantics, such as art, biology, geography, history, and warfare. The dataset is split as 3 parts following \cite{feng12014cross,DBLP:conf/ijcai/PengHQ16}, where the training set has 2,173 pairs, the testing set has 462 pairs, and the validation set has 231 pairs. 


		{\textbf {NUS-WIDE-10k dataset}} \cite{feng12014cross} is a subset selected from a large-scale image/tag dataset NUS-WIDE \cite{NUSWIDE}. NUS-WIDE dataset has totally 270,000 images with corresponding text tags, categorized into 81 classes. Different from Wikipedia dataset, in NUS-WIDE dataset the text modality refers to tags, instead of textual descriptions as paragraphs or sentences.  NUS-WIDE-10k is constructed by randomly selecting 1,000 image/tag pairs from each of the 10 largest classes, where each image/tag pair only belongs to one of them. The totally 10,000 pairs are randomly split into 3 parts: the training set has 8,000 pairs, the testing set has 1,000 pairs, and the validation set has 1,000 pairs. Note that the pairs in the 3 parts are all evenly selected from the 10 classes.

		{\textbf {Pascal Sentences dataset}} \cite{PascalSentence} is selected from 2008 PASCAL development kit, which contains 1,000 images with corresponding text descriptions as 5 sentences. These image/text pairs can be evenly classified into 20 classes. Similar split strategy with NUS-WIDE-10k dataset is adopted for Pascal Sentences dataset, where the training set has 800 pairs, the testing set has 100 pairs, and the validation set has 100 pairs, all of which are evenly from the 20 classes.
		
\begin{figure}[t]
			\begin{center}
				\includegraphics[width=0.95\linewidth]{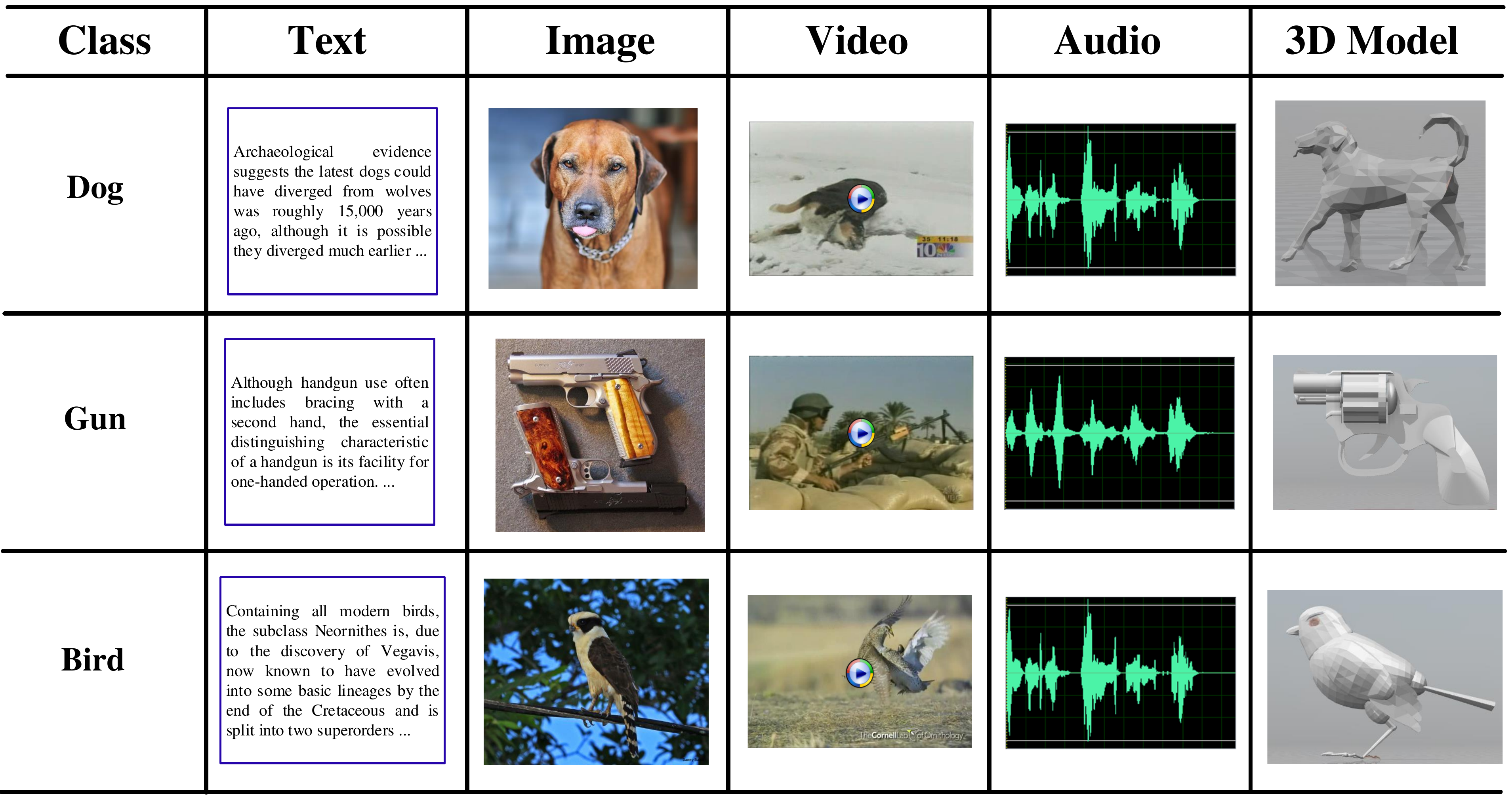}
			\end{center}
			\caption{Examples of XMedia dataset from 3 classes: dog, gun and bird.}
			\label{fig:exmPub1}
		\end{figure}

		{\textbf{XMedia dataset}} \cite{DBLP:journals/corr/PengHZ17a} is the first publicly available cross-modal dataset with up to 5 modalities (text, image, video, audio and 3D model), for comprehensive evaluation of cross-modal retrieval \cite{ZhaiTCSVT2014JRL,PengHypergraph2015}. There are totally 20 classes in XMedia dataset, which are specific objects such as insect, bird, wind, dog, and elephant. There are 250 texts, 250 images, 25 videos, 50 audio clips and 25 3D models for each class, and the total number of instances is 12,000. Some examples of the dataset are shown in Figure \ref{fig:exmPub1}. Similar to other datasets, we split XMedia dataset as 3 parts: the training set has 9,600 instances, the testing set has 1,200 instances, and the validation set has 1,200 instances. Note that in each set, the instances are evenly from the 5 modalities for ensuring the variety.
		
		Although all the datasets are split into training, testing and validation sets, it should be noted that not all compared methods take validation set as input. Only CMDN, Corr-AE, Bimodal AE, and Multimodal DBN actually use the validation set as input. For the other compared methods including our MHTN, the validation set is not used in both training and testing stages.
		
		\subsection{Retrieval Tasks and Evaluation Metrics}
		
		In the experiments, bi-modal retrieval is conducted for evaluation, which means that the retrieval is performed between two different modalities. For instance, on Wikipedia dataset we retrieve the texts by image queries (denoted as Image$\rightarrow$Text) and vice versa, and on XMedia dataset the retrieval will be performed between each pair of modalities. As for the retrieval process, taking Image$\rightarrow$Text as an example, we first obtain the common representation for all the images and texts in testing set with all compared methods and MHTN. Then we select each image in testing set as the query, compute the similarity between the query and every text in testing set by cosine distance, and finally get the similarity ranking list.

		The metric adopted for evaluating the retrieval results is mean average precision (MAP) score, which is the mean value of average precision (AP) scores of all queries. For each query, the AP score can be computed as follows: 
		\begin{align}
			AP = \frac 1 R \sum_{k=1}^n \frac {R_k} k \times rel_k
		\end{align}
		where $R$ denotes the number of relevant items in the testing set, $R_k$ is the number of relevant items in top $k$ results in the ranking list, and $n$ is the total number of instances in testing set. $rel_k = 1$ if the query and $k$-th result in the ranking list belong to the same class (i.e., they are relevant), and $rel_k = 0$  otherwise.  \emph{All the retrieval results} will be considered for the computation of MAP score. We also adopt precision-recall (PR) curve for more comprehensive evaluation, which shows the search precision at all recall levels.

		\renewcommand{\multirowsetup}{\centering}  
		\begin{table*}[tbp]
			\tiny
			\centering
			\caption{The MAP scores of cross-modal retrieval of our MHTN compared with state-of-the-art methods.}
			\label{table:Results}
			\begin{tabular}{c|c|c|c|c|c|c|c|c|c|c|c|c|c}
				\hline
				\multirow{2}{*} {Dataset} 
				& \multirow{2}{*} {Task}
				& \multirow{2}{*} {CCA} 
				& \multirow{2}{*} {CFA} 
				& KCCA
				& KCCA
				& Bimodal 
				& Multimodal 
				& \multirow{2}{*} {Corr-AE} 
				& \multirow{2}{*} {JRL} 
				& \multirow{2}{*} {LGCFL } 
				& \multirow{2}{*} {CMDN } 
				& \multirow{2}{*} {Deep-SM } 
				& \multirow{2}{*} {Our MHTN } 
				\\  & & & & (Poly) & (Gaussian) & AE & DBN & & & & & &  \\
				\hline \hline
				\multirow{3}{*}{\begin{tabular}{c}Wikipedia\end{tabular}} & Image$\rightarrow$Text & 0.176  & 0.330  & 0.230  & 0.357  & 0.301  & 0.204  & 0.373  & 0.408  & 0.416  & 0.409  & 0.458  & 0.514  \\
				& Text$\rightarrow$Image & 0.178  & 0.306  & 0.224  & 0.328  & 0.267  & 0.145  & 0.357  & 0.353  & 0.360  & 0.364  & 0.345  & 0.444  \\
				& \textbf{Average} & \textbf{0.177}  & \textbf{0.318}  & \textbf{0.227}  & \textbf{0.343}  & \textbf{0.284}  & \textbf{0.175}  & \textbf{0.365}  & \textbf{0.381}  & \textbf{0.388}  & \textbf{0.387}  & \textbf{0.402}  & \textbf{0.479}  \\
				\hline \hline
				\multirow{3}{*}{\begin{tabular}{c}NUS-WIDE-10k\end{tabular}} & Image$\rightarrow$Text & 0.159  & 0.299  & 0.129  & 0.295  & 0.234  & 0.178  & 0.306  & 0.410  & 0.408  & 0.410  & 0.389  & 0.520  \\
				& Text$\rightarrow$Image & 0.189  & 0.301  & 0.157  & 0.162  & 0.376  & 0.144  & 0.340  & 0.444  & 0.374  & 0.450  & 0.496  & 0.534  \\
				& \textbf{Average} & \textbf{0.174}  & \textbf{0.300}  & \textbf{0.143}  & \textbf{0.229}  & \textbf{0.305}  & \textbf{0.161}  & \textbf{0.323}  & \textbf{0.427}  & \textbf{0.391}  & \textbf{0.430}  & \textbf{0.443}  & \textbf{0.527}  \\
				\hline \hline
				
				\multirow{3}{*}{\begin{tabular}{c}Pascal Sentences \end{tabular}} & Image$\rightarrow$Text & 0.110  & 0.341  & 0.271  & 0.312  & 0.404  & 0.438  & 0.411  & 0.416  & 0.381  & 0.458  & 0.440  & 0.496  \\
				& Text$\rightarrow$Image & 0.116  & 0.308  & 0.280  & 0.329  & 0.447  & 0.363  & 0.475  & 0.377  & 0.435  & 0.444  & 0.414  & 0.500  \\
				& \textbf{Average} & \textbf{0.113}  & \textbf{0.325}  & \textbf{0.276}  & \textbf{0.321}  & \textbf{0.426}  & \textbf{0.401}  & \textbf{0.443}  & \textbf{0.397}  & \textbf{0.408}  & \textbf{0.451}  & \textbf{0.427}  & \textbf{0.498}  \\
				\hline 	\hline 
				\multirow{21}{*}{\begin{tabular}{c}XMedia\end{tabular}} & Image$\rightarrow$Text & 0.257  & 0.292  & 0.324  & 0.447  & 0.598  & 0.093  & 0.450  & 0.770  & 0.744  & 0.794  & 0.822  & 0.853  \\
				& Image$\rightarrow$Video & 0.179  & 0.451  & 0.215  & 0.421  & 0.301  & 0.187  & 0.354  & 0.653  & 0.667  & 0.617  & 0.727  & 0.753  \\
				& Image$\rightarrow$Audio & 0.159  & 0.375  & 0.115  & 0.342  & 0.111  & 0.111  & 0.141  & 0.472  & 0.110  & 0.376  & 0.696  & 0.730  \\
				& Image$\rightarrow$3D & 0.256  & 0.443  & 0.266  & 0.284  & 0.371  & 0.215  & 0.344  & 0.569  & 0.605  & 0.540  & 0.757  & 0.803  \\
				& Text$\rightarrow$Image & 0.341  & 0.283  & 0.340  & 0.590  & 0.642  & 0.120  & 0.437  & 0.800  & 0.804  & 0.805  & 0.807  & 0.843  \\
				& Text$\rightarrow$Video & 0.203  & 0.266  & 0.198  & 0.335  & 0.339  & 0.252  & 0.445  & 0.682  & 0.708  & 0.655  & 0.647  & 0.696  \\
				& Text$\rightarrow$Audio & 0.240  & 0.211  & 0.129  & 0.300  & 0.103  & 0.103  & 0.145  & 0.515  & 0.163  & 0.354  & 0.622  & 0.689  \\
				& Text$\rightarrow$3D & 0.323  & 0.264  & 0.260  & 0.250  & 0.370  & 0.225  & 0.340  & 0.675  & 0.641  & 0.554  & 0.647  & 0.733  \\
				& Video$\rightarrow$Image & 0.231  & 0.339  & 0.165  & 0.297  & 0.416  & 0.091  & 0.308  & 0.602  & 0.630  & 0.578  & 0.700  & 0.725  \\
				& Video$\rightarrow$Text & 0.289  & 0.178  & 0.118  & 0.229  & 0.439  & 0.079  & 0.361  & 0.613  & 0.664  & 0.609  & 0.634  & 0.699  \\
				& Video$\rightarrow$Audio & 0.172  & 0.213  & 0.094  & 0.213  & 0.110  & 0.098  & 0.115  & 0.381  & 0.091  & 0.349  & 0.572  & 0.632  \\
				& Video$\rightarrow$3D & 0.276  & 0.258  & 0.120  & 0.258  & 0.277  & 0.104  & 0.263  & 0.474  & 0.493  & 0.516  & 0.603  & 0.659  \\
				& Audio$\rightarrow$Image & 0.076  & 0.314  & 0.087  & 0.212  & 0.088  & 0.073  & 0.078  & 0.424  & 0.079  & 0.365  & 0.668  & 0.694  \\
				& Audio$\rightarrow$Text & 0.092  & 0.223  & 0.089  & 0.212  & 0.089  & 0.069  & 0.083  & 0.486  & 0.084  & 0.375  & 0.607  & 0.667  \\
				& Audio$\rightarrow$Video & 0.144  & 0.305  & 0.108  & 0.284  & 0.147  & 0.149  & 0.131  & 0.418  & 0.121  & 0.405  & 0.566  & 0.599  \\
				& Audio$\rightarrow$3D & 0.173  & 0.235  & 0.145  & 0.204  & 0.156  & 0.137  & 0.148  & 0.430  & 0.116  & 0.424  & 0.574  & 0.614  \\
				& 3D$\rightarrow$Image & 0.123  & 0.326  & 0.189  & 0.187  & 0.261  & 0.063  & 0.196  & 0.452  & 0.258  & 0.473  & 0.696  & 0.697  \\
				& 3D$\rightarrow$Text & 0.195  & 0.203  & 0.160  & 0.151  & 0.283  & 0.063  & 0.243  & 0.619  & 0.309  & 0.518  & 0.649  & 0.678  \\
				& 3D$\rightarrow$Video & 0.148  & 0.311  & 0.135  & 0.184  & 0.146  & 0.145  & 0.168  & 0.502  & 0.334  & 0.445  & 0.570  & 0.589  \\
				& 3D$\rightarrow$Audio & 0.113  & 0.231  & 0.125  & 0.209  & 0.095  & 0.079  & 0.098  & 0.427  & 0.113  & 0.363  & 0.579  & 0.607  \\
				& \textbf{Average} & \textbf{0.200}   & \textbf{0.286}   & \textbf{0.169}   & \textbf{0.280}   & \textbf{0.267}   & \textbf{0.123}   & \textbf{0.242}   & \textbf{0.548}   & \textbf{0.387}   & \textbf{0.506}   & \textbf{0.657}   & \textbf{0.698}  \\
				\hline
				
			\end{tabular}%
		\end{table*}%

		\subsection{Compared Methods and Input Settings}
		
		In the experiments, we compare our proposed MHTN approach with totally 10 state-of-the-art methods, namely 
		CCA \cite{HotelingBiometrika36RelationBetweenTwoVariates}, CFA \cite{LiMM03CFA}, KCCA (with Gaussian and polynomial kernel) \cite{DBLP:journals/neco/HardoonSS04}, Bimodal AE \cite{ngiam32011multimodal}, Multimodal DBN \cite{srivastava2012learning}, Corr-AE \cite{feng12014cross}, JRL \cite{ZhaiTCSVT2014JRL}, LGCFL \cite{DBLP:journals/tmm/KangXLXP15}, CMDN \cite{DBLP:conf/ijcai/PengHQ16} and Deep-SM \cite{DBLP:journals/tcyb/WeiZLWLZY17}. 
		Traditional methods include CCA, CFA, KCCA, JRL and LGCFL, while DNN-based methods include Bimodal-AE, Multimodal DBN, Corr-AE, CMDN and Deep-SM.
		It should be noted that JRL and our proposed MHTN can jointly learn common representation for \emph{all modalities} simultaneously, while the other compared methods can only learn common representation for \emph{two modalities} at a time.

		For \textbf{image}, the architectures of Deep-SM and our MHTN approach have CNNs built-in (AlexNet is adopted in the experiments), so they take the original image pixels as input, while all the other methods take extracted feature vectors as input. So for all the methods except MHTN and Deep-SM, we use the same AlexNet pre-trained on ImageNet which is  further fine-tuned with the images in each dataset to convergence, and extract the output of 4,096 dimensions from fc7 layer as the image features. 
		
		For text, video, audio, and 3D model, exactly the same features are used for all compared methods and our MHTN: 
		For \textbf{text}, exactly following \cite{feng12014cross,DBLP:conf/ijcai/PengHQ16}, the 3,000-dimensional BoW features are adopted for Wikipedia dataset, and the 1,000-dimensional BoW features are adopted for NUS-WIDE-10k and Pascal Sentences datasets. On XMedia dataset we take 3,000-dimensional BoW text features, which are the same with Wikipedia dataset. 
		For \textbf{video}, we use C3D model \cite{tran2015learning} pre-trained on Sports1M \cite{karpathy2014large} to extract the output of 4,096 dimensions from fc7 layer as the video features. 
		For \textbf{audio}, audio clips are represented by the 78-dimensional features, which are extracted by jAudio \cite{mckay2005jaudio} using its default setting.
		For \textbf{3D model}, the models are represented by the concatenated 4,700-dimensional vectors of a LightField descriptor set \cite{chen2003visual}.
		
		\begin{figure}[t]
			\centering
			\begin{minipage}[c]{\linewidth}
				\centering
				\includegraphics[width=\textwidth]{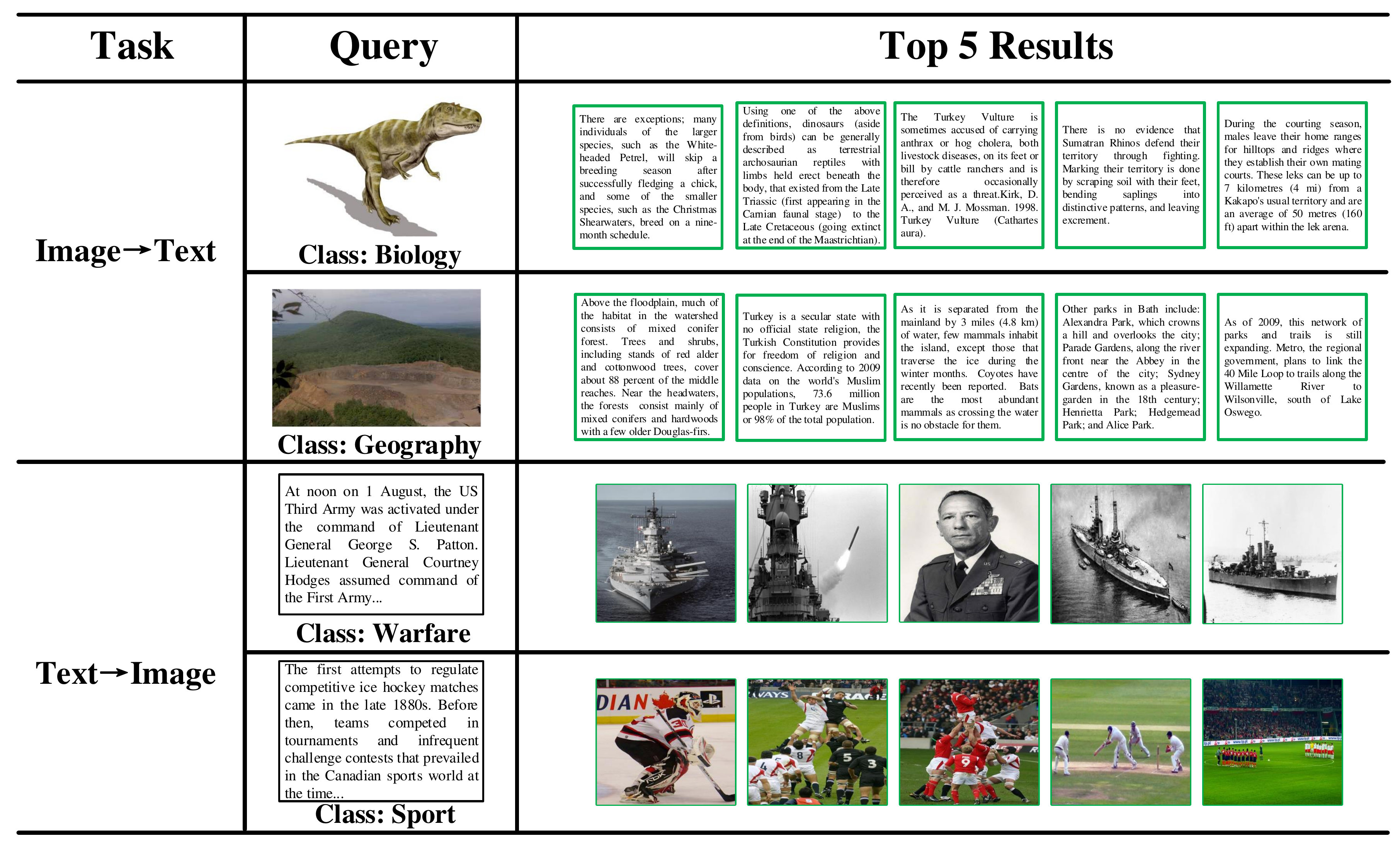}
			\end{minipage}%
			\setlength{\abovecaptionskip}{0.2cm}
			\caption{Some examples of retrieval results on Wikipedia dataset, where all the top 5 retrieval results are correct.}\label{fig:WikiRes}
		\end{figure}

		\begin{figure}[t]
			\centering
			\begin{minipage}[c]{\linewidth}
				\centering
				\includegraphics[width=\textwidth]{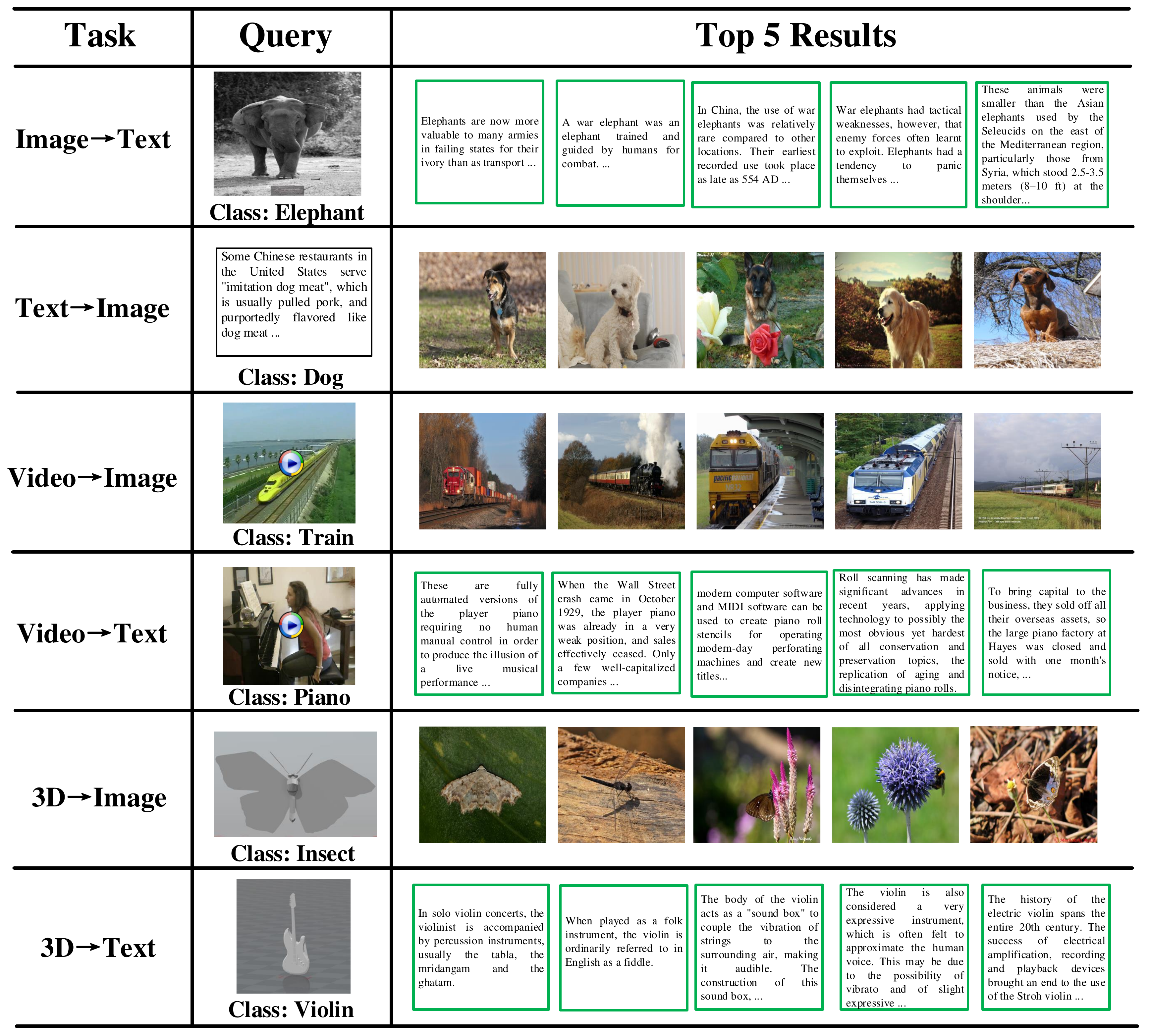}
			\end{minipage}%
			\setlength{\abovecaptionskip}{0.2cm}
			\caption{Some examples of retrieval results on XMedia dataset, where all the top 5 retrieval results are correct.}\label{fig:XMediaRes}
		\end{figure}
		
		\subsection{Experimental Results}
\begin{figure}[tp]
			\begin{center}
				
				\subfigure[The PR curves on Wikipedia dataset.]{
					\begin{minipage}[c]{0.5\textwidth}
						\includegraphics[width=0.46\textwidth]{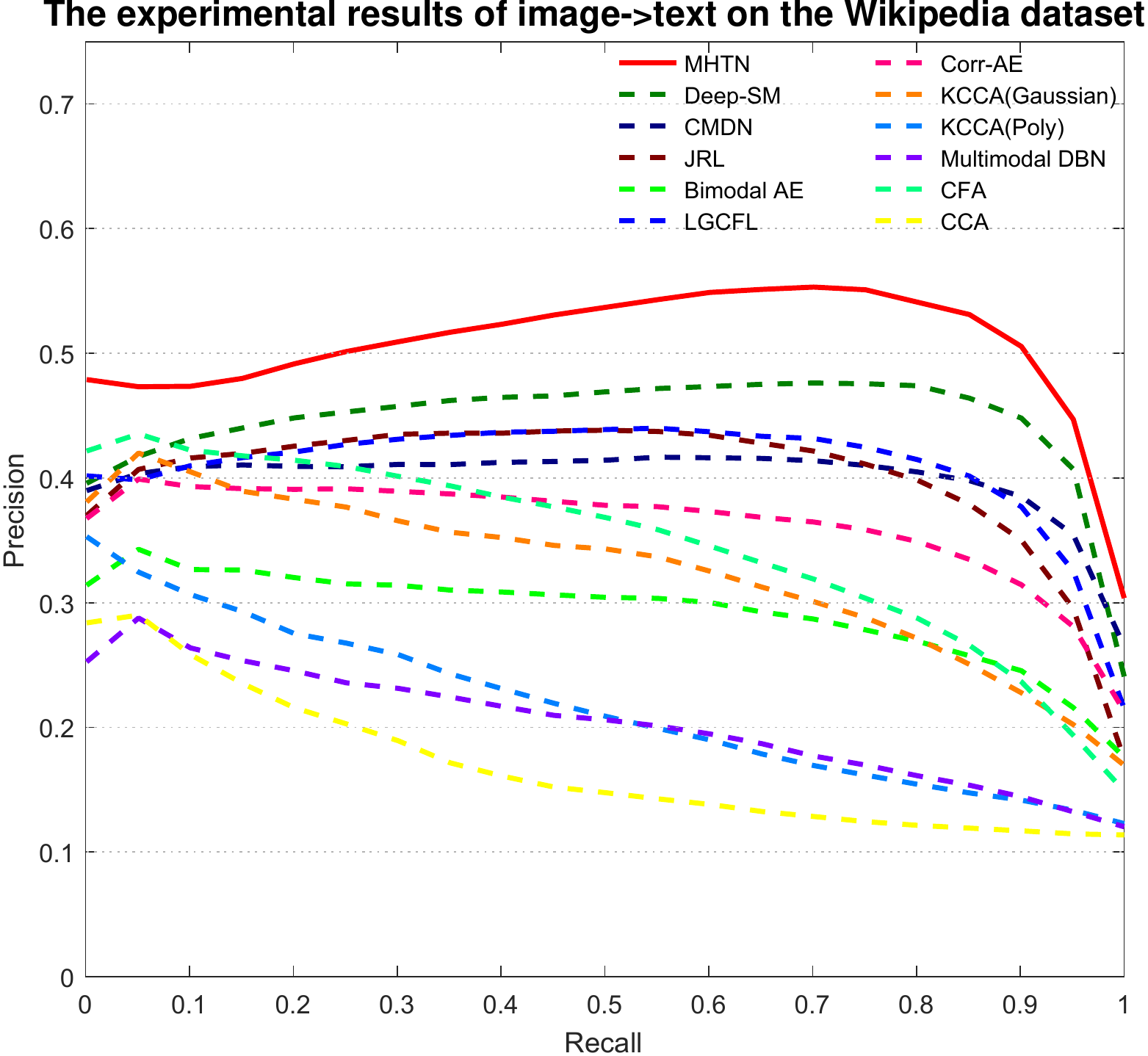}
						\includegraphics[width=0.46\textwidth]{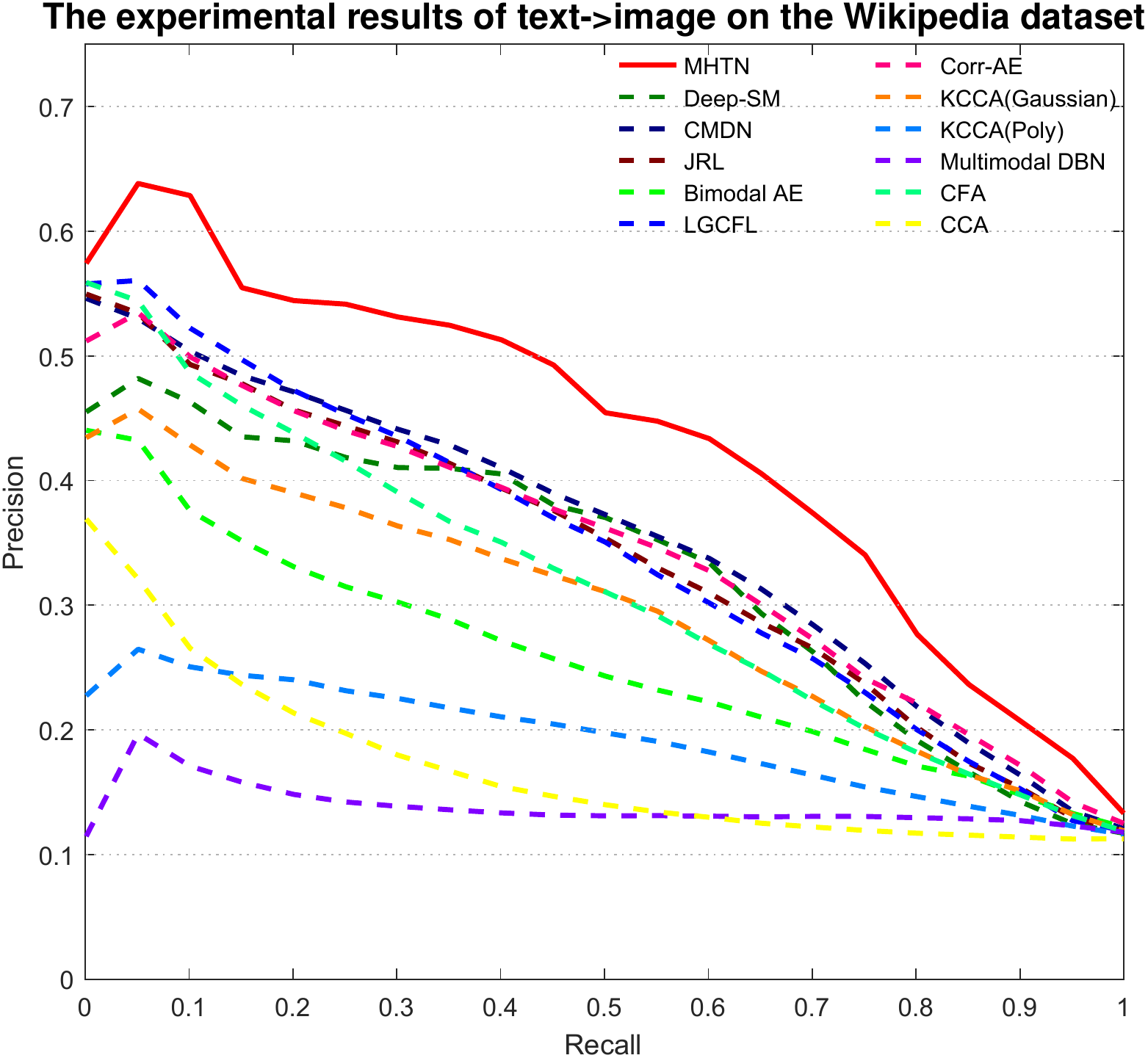}
						\vspace{2mm}
					\end{minipage}
				}
				
				\subfigure[The PR curves on NUS-WIDE-10k dataset.]{
					\begin{minipage}[c]{0.5\textwidth}
						\includegraphics[width=0.47\textwidth]{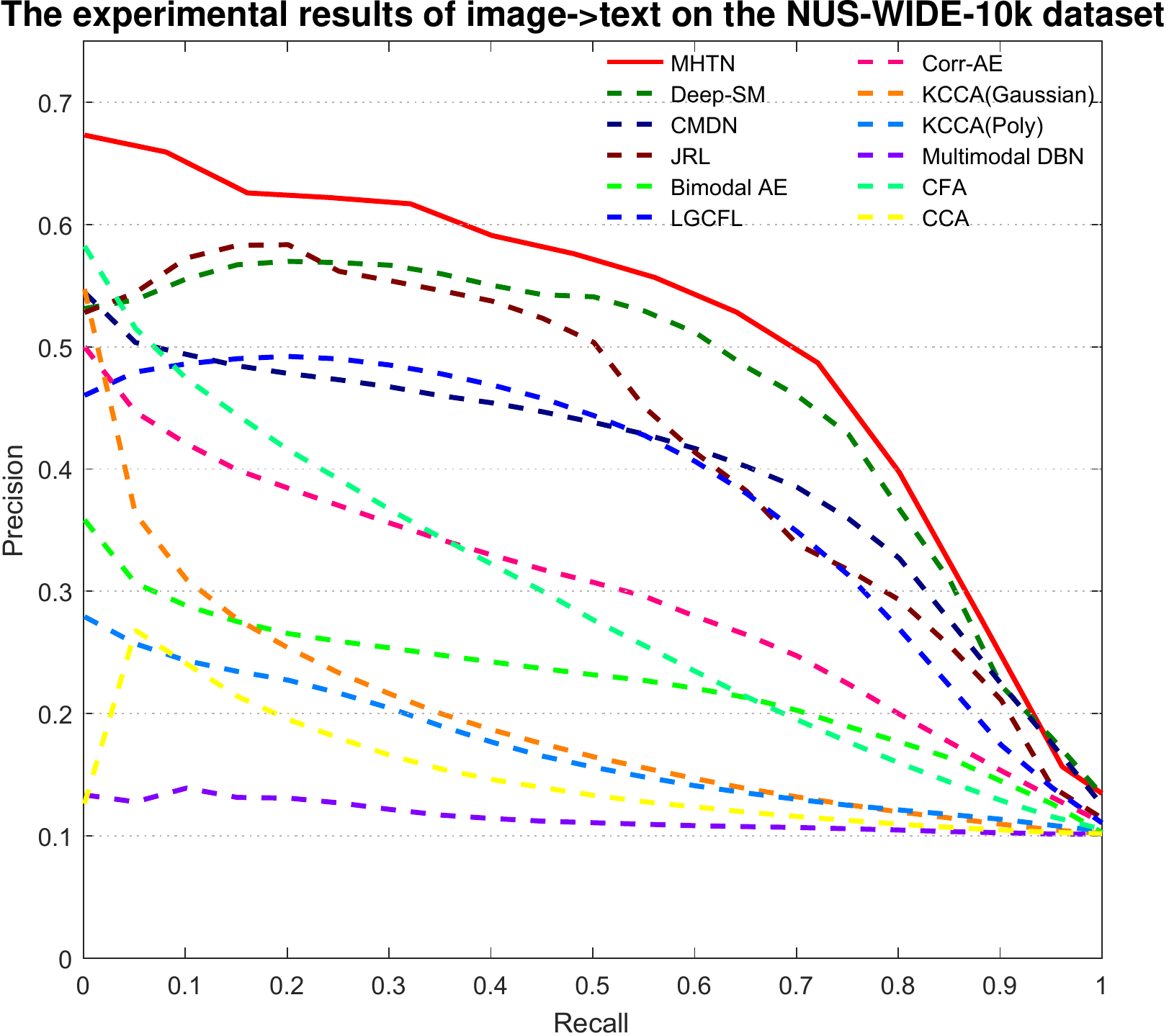}
						\includegraphics[width=0.47\textwidth]{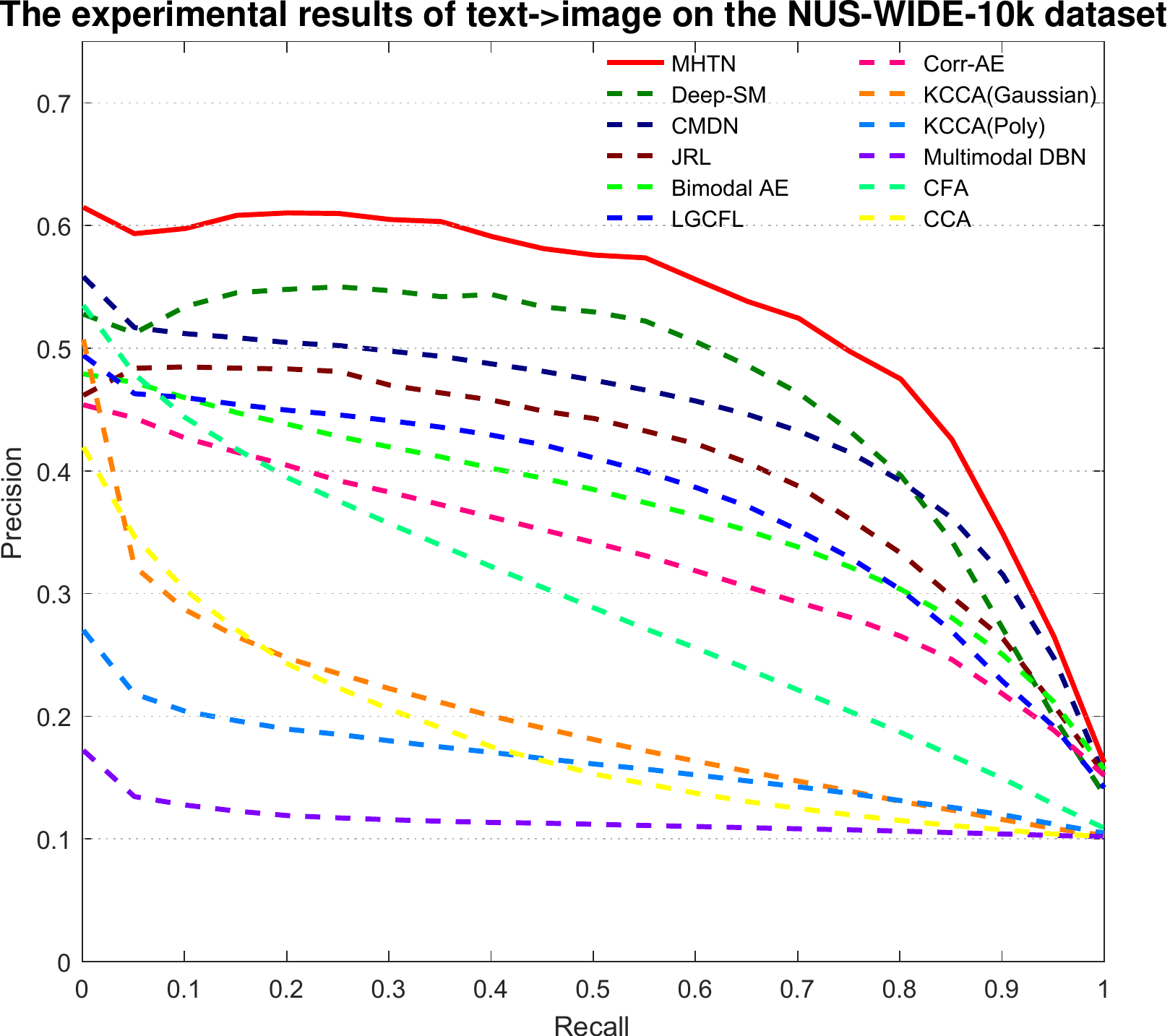}
						\vspace{2mm}
					\end{minipage}
				}
				
				\subfigure[The PR curves on Pascal Sentences dataset.]{
					\begin{minipage}[c]{0.5\textwidth}
						\includegraphics[width=0.48\textwidth]{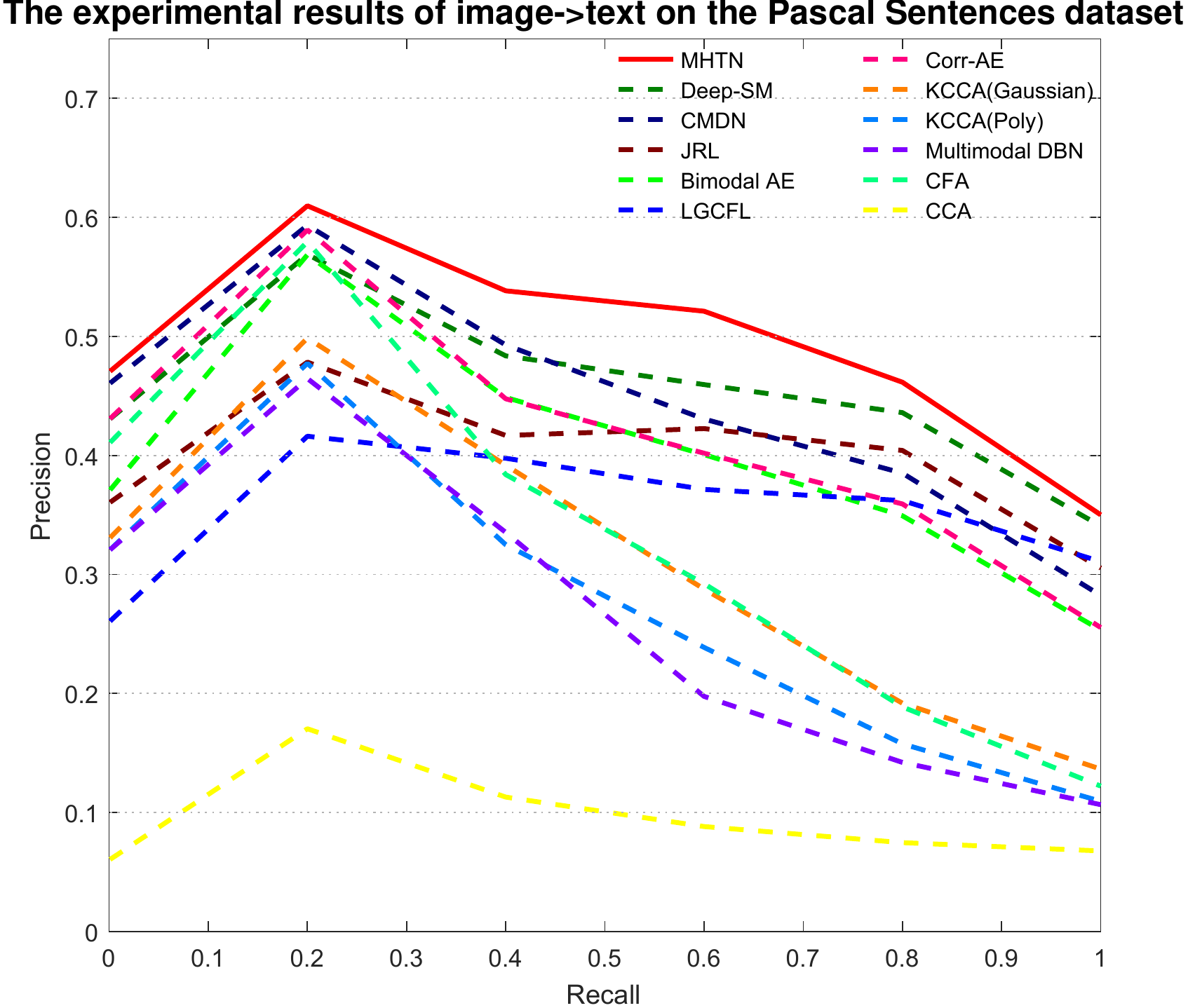}
						\includegraphics[width=0.48\textwidth]{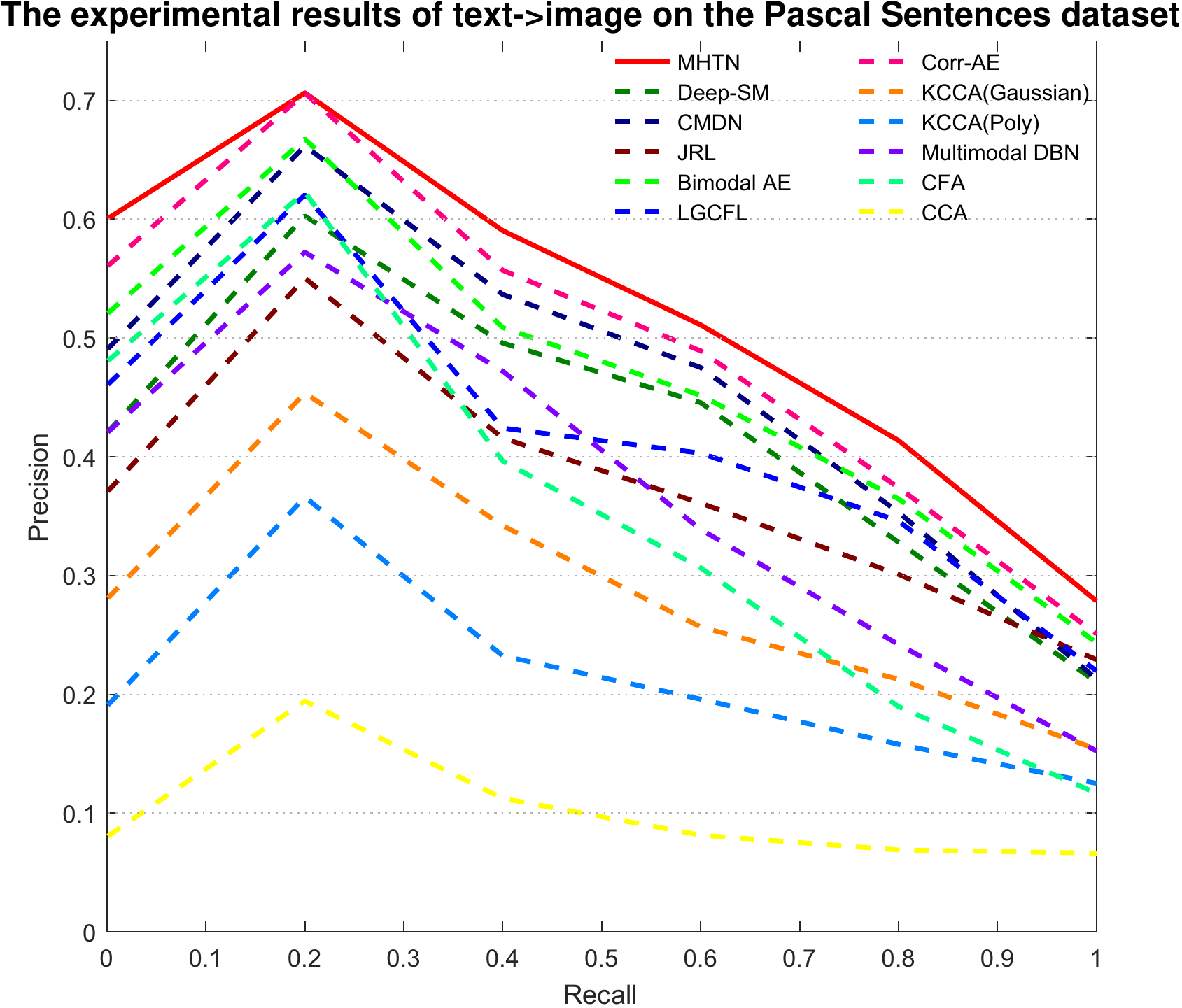}
						\vspace{2mm}
					\end{minipage}
				}
				
			\end{center}
			\caption{The PR curves on Wikipedia, NUS-WIDE-10k and Pascal Sentences datasets.}
			\label{fig:PR_All}
		\end{figure}

		Table \ref{table:Results} shows the experimental results of MAP scores of our MHTN approach as well as state-of-the-art methods. On all the 4 datasets, our proposed MHTN achieves the highest MAP scores on all retrieval tasks. 
		On Wikipedia dataset, the highest average MAP score of compared methods is obtained by Deep-SM, and an inspiring accuracy improvement is obtained by MHTN from 0.402 to 0.479. Figure \ref{fig:WikiRes} shows some retrieval results of our MHTN approach on Wikipedia dataset.
		On NUS-WIDE-10k dataset, MHTN keeps its advantage and achieves the highest average MAP score of 0.527. 
		On Pascal Sentences dataset, we can see that CMDN obtains the highest average MAP score of compared methods, but MHTN still achieves a clear advantage to be 0.498. 
		On XMedia dataset, the performance trends among compared methods differ from the above 3 datasets. For example, the accuracy of DNN-based methods such as Multimodal DBN, Bimodal AE and Corr-AE is clearly lower than traditional methods such as LGCFL and JRL. The reason is that these methods are mainly based on pairwise correlation, but the numbers of instances for some modalities are small such as totally 500 3D models, which makes it hard to capture the semantic consistency only by pairwise correlation. However, our proposed MHTN remains the best accuracy, which shows its effectiveness and generality. Figure \ref{fig:XMediaRes} shows some retrieval results of our MHTN method on XMedia dataset.
		Figure \ref{fig:PR_All} shows the PR curves of all retrieval tasks on Wikipedia, NUS-WIDE-10k and Pascal Sentences datasets, where our proposed MHTN keeps the highest precisions at all recall levels. 
		
		For fair comparison, all the compared methods except Deep-SM take CNN image features extracted from AlexNet \emph{pre-trained on ImageNet} and fine-tuned with the images in the target domain, and Deep-SM takes the fine-tuned AlexNet as its component for image. In fact, this can be seen as straight-forward single-modal knowledge transfer from images in source domain only to images in cross-modal target domain. 
		However, in this way, the knowledge transfer across different modalities is not jointly involved, which results in inadequate transfer and limited accuracy.
		Our proposed MHTN achieves a clear advantage on all datasets. On the one hand, it can jointly transfer knowledge from a single modality in source domain to multiple modalities in cross-modal target domain, which can distill the modal-independent knowledge to enrich training information and avoid overfitting. On the other hand, the modal-adversarial training strategy can ensure the cross-modal semantic consistency in the hybrid transfer process, and further  improve the accuracy of retrieval.

		\begin{table*}[htbp]
			\scriptsize
			\centering
			\caption{The MAP scores of experiments to show the impacts of MHTN's components.}
			\label{table:Baseline}
			\begin{tabular}{c|c|c|c|c|c|c|c}
				\hline
				\multirow{2}{*} {Dataset} 
				& \multirow{2}{*} {Task}
				& MHTN 
				& MHTN 
				& MHTN
				& MHTN
				& MHTN 
				& MHTN
				\\  & & (NoSource) & (NoSLnet) & (NoAdver) & (NoSDS) & (Full) & (VGG19)\\
				\hline \hline
				\multirow{3}{*}{\begin{tabular}{c}Wikipedia\end{tabular}} & Image$\rightarrow$Text & 0.478  & 0.483  & 0.508  & 0.494  & 0.514  & 0.541 \\
				& Text$\rightarrow$Image & 0.414  & 0.422  & 0.432  & 0.420  & 0.444  & 0.461 \\
				& \textbf{Average} & \textbf{0.446}  & \textbf{0.453}  & \textbf{0.470}  & \textbf{0.457}  & \textbf{0.479}  & \textbf{0.501} \\
				\hline \hline
				
				\multirow{3}{*}{\begin{tabular}{c}NUS-WIDE-10k\end{tabular}} & Image$\rightarrow$Text & 0.489  & 0.488  & 0.518  & 0.498  & 0.520  & 0.552 \\
				& Text$\rightarrow$Image & 0.515  & 0.442  & 0.516  & 0.518  & 0.534  & 0.541 \\
				& \textbf{Average} & \textbf{0.502}  & \textbf{0.465}  & \textbf{0.517}  & \textbf{0.508}  & \textbf{0.527}  & \textbf{0.547} \\
				\hline \hline
				
				\multirow{3}{*}{\begin{tabular}{c}Pascal Sentences\end{tabular}} & Image$\rightarrow$Text & 0.449  & 0.446  & 0.467  & 0.473  & 0.496  & 0.572 \\
				& Text$\rightarrow$Image & 0.464  & 0.467  & 0.477  & 0.493  & 0.500  & 0.561 \\
				& \textbf{Average} & \textbf{0.457}  & \textbf{0.457}  & \textbf{0.472}  & \textbf{0.483}  & \textbf{0.498}  & \textbf{0.567} \\
				\hline 	\hline
				\multirow{21}{*}{\begin{tabular}{c}XMedia\end{tabular}} & Image$\rightarrow$Text & 0.820  & 0.836  & 0.853  & 0.835  & 0.853  & 0.892 \\
				& Image$\rightarrow$Video & 0.726   & 0.741   & 0.741   & 0.723   & 0.753  & 0.781 \\
				& Image$\rightarrow$Audio & 0.727   & 0.686   & 0.753   & 0.719   & 0.730  & 0.751 \\
				& Image$\rightarrow$3D & 0.765   & 0.748   & 0.717   & 0.773   & 0.803  & 0.802 \\
				& Text$\rightarrow$Image & 0.813   & 0.823   & 0.838   & 0.821   & 0.843  & 0.880 \\
				& Text$\rightarrow$Video & 0.667   & 0.670   & 0.691   & 0.670   & 0.696  & 0.734 \\
				& Text$\rightarrow$Audio & 0.666   & 0.626   & 0.694   & 0.671   & 0.689  & 0.692 \\
				& Text$\rightarrow$3D & 0.681   & 0.669   & 0.659   & 0.712   & 0.733  & 0.717 \\
				& Video$\rightarrow$Image & 0.675   & 0.705   & 0.711   & 0.708   & 0.725  & 0.742 \\
				& Video$\rightarrow$Text & 0.623   & 0.654   & 0.688   & 0.673   & 0.699  & 0.704 \\
				& Video$\rightarrow$Audio & 0.578   & 0.550   & 0.622   & 0.613   & 0.632  & 0.609 \\
				& Video$\rightarrow$3D & 0.591   & 0.615   & 0.602   & 0.636   & 0.659  & 0.628 \\
				& Audio$\rightarrow$Image & 0.691   & 0.698   & 0.692   & 0.712   & 0.694  & 0.726 \\
				& Audio$\rightarrow$Text & 0.645   & 0.643   & 0.667   & 0.675   & 0.667  & 0.685 \\
				& Audio$\rightarrow$Video & 0.584   & 0.587   & 0.593   & 0.597   & 0.599  & 0.618 \\
				& Audio$\rightarrow$3D & 0.596   & 0.570   & 0.577   & 0.596   & 0.614  & 0.607 \\
				& 3D$\rightarrow$Image & 0.684   & 0.690   & 0.671   & 0.661   & 0.697  & 0.746 \\
				& 3D$\rightarrow$Text & 0.648   & 0.646   & 0.654   & 0.648   & 0.678  & 0.716 \\
				& 3D$\rightarrow$Video & 0.535   & 0.561   & 0.559   & 0.548   & 0.589  & 0.607 \\
				& 3D$\rightarrow$Audio & 0.603   & 0.548   & 0.601   & 0.571   & 0.607  & 0.623 \\
				& \textbf{Average} & \textbf{0.666}  & \textbf{0.663}  & \textbf{0.679}  & \textbf{0.678}  & \textbf{0.698}  & \textbf{0.713} \\
				\hline
				
			\end{tabular}%
		\end{table*}%

		\subsection{Impacts of Components in Our MHTN}
		
		Because MHTN consists of multiple components, we further conduct experiments for evaluating the impacts of them. All of the experimental results are shown in Table \ref{table:Baseline}, which are introduced and analyzed as follows:
		
		\begin{itemize}
			\item {\bf The impact of source image pathway}. 
			The source image path way (the top pathway in Figure \ref{fig:network}) is a key component of hybrid transfer learning, which acts as the source for knowledge transfer. In Table \ref{table:Baseline}, MHTN (NoSource)
			means that we remove the source image pathway, and the other components remain the same.  MHTN (Full) means the complete MHTN. By comparing the results of MHTN (NoSource) and MHTN (Full) in Table \ref{table:Baseline}, we can see that the source image pathway 
			stably improves the retrieval accuracy. This is because single-modal datasets contain not only modal-specific information, but also rich modal-independent semantic knowledge that can be jointly shared across different modalities, which can provide considerable supplementary information for cross-modal retrieval.

			\item {\bf The impact of modal-adversarial semantic learning subnetwork}. 
			Modal-adversarial semantic learning subnetwork is designed to make the process of knowledge transfer further adapted to cross-modal retrieval task. In Table \ref{table:Baseline}, MHTN (NoSLnet) means that we remove the modal-adversarial semantic learning subnetwork, and directly train a classifier (as fc10 and softmax layer in Figure \ref{fig:subnet2} ) with the specific representation of each modality, which generates probability vectors for retrieval. By comparing MHTN (NoSLnet) and MHTN (Full) in Table \ref{table:Baseline}. We can see that MHTN (Full) achieves higher results on all 4 datasets, which shows that this subnetwork can preserve the inherent cross-modal semantic consistency in target domain and improve the retrieval accuracy.
			
			\item {\bf The impact of modal-adversarial training strategy}. 
			Modal-adversarial training strategy is used to explicitly reduce ``heterogeneity gap". In Table \ref{table:Baseline}, MHTN (NoAdver) means that we remove the modal-adversarial consistency learning part, which is exactly  the same with with our previous conference paper \cite{DBLP:journals/corr/HuangPY17} on Wikipedia, NUS-WIDE-10k and Pascal Sentences datasets. By comparing MHTN (NoAdver) and MHTN (Full), we can see the modal-adversarial improves the retrieval accuracy by making the common representation \emph{discriminative for semantics} but \emph{indiscriminative for modalities}, which effectively enhance the cross-modal semantic consistency.
			
			\item {\bf The impact of source domain supervision loss}. 
			Because the classes of source domain and target domain are different, this experiment aims to verify if the fine-tuning of source image pathway can improve the retrieval accuracy. In Table \ref{table:Baseline}, MHTN (NoSDS) means that we remove the source domain supervision loss ($Loss_{SDS}$). By comparing  MHTN (NoSDS) and MHTN (Full), we can see that although the classes are different of the two domains, the supervision information in source domain still contains general knowledge, which can be shared by cross-modal target domain and improve retrieval accuracy.
			
			\item {\bf The impact of different CNN structures}. 
			We further conduct an experiment where the AlexNex in source image pathway and target image pathway are replaced by VGG19 \cite{DBLP:journals/corr/SimonyanZ14a}, which is denoted as MHTN (VGG19) in Table \ref{table:Baseline}. Except the difference of CNN structure, the other components keep the same. It can be seen that VGG19 can improve the overall retrieval accuracy on 4 datasets, especially on Pascal Sentences dataset. Note that on XMedia dataset, we only modify the image pathways, but most retrieval tasks even without image (such as Text$\rightarrow$Video and 3D$\rightarrow$Text) can also benefit a lot. It shows the effectiveness and generality of MHTN for transferring knowledge to all modalities.

		\end{itemize}
		
		

		\section{Conclusion}
		
		This paper has proposed modal-adversarial hybrid transfer network (MHTN), which aims to transfer knowledge from \emph{single-modal source domain} to \emph{cross-modal target domain} for promoting cross-modal retrieval. It is an end-to-end architecture with two subnetworks: 
		\emph{Modal-sharing knowledge transfer subnetwork} is proposed to jointly transfer knowledge from a large-scale single-modal dataset in source domain to all modalities in target domain with a star network structure, which distills modal-independent supplementary knowledge for promoting cross-modal common representation learning.
		\emph{Modal-adversarial semantic learning subnetwork} is proposed to construct an adversarial training mechanism between common representation generator and modality discriminator, making the common representation \emph{discriminative for semantics} but \emph{indiscriminative for modalities} to enhance cross-modal semantic consistency during transfer process.
		Comprehensive experiments on 4 datasets show the effectiveness and generality of MHTN, including challenging XMedia dataset with 5 modalities (text, image, video, audio, and 3D model).
		
		In the future, we intend to improve the work in the following two aspects: First, we will apply the architecture of MHTN to other applications involving cross-modal data such as image caption. Second, we will attempt to address the problem of hybrid knowledge transfer under unsupervised setting, which aims to bring stronger flexibility to MHTN.

		
		%
		\bibliographystyle{IEEEtran}
		\bibliography{cite}

		\appendices

	\end{document}